\documentstyle[aps,twocolumn]{revtex}
\input psfig

\begin{document}

\twocolumn[\hsize\textwidth\columnwidth\hsize\csname
@twocolumnfalse\endcsname

\draft

\title{Statistical Mechanics of Cracks: Thermodynamic Limit, Fluctuations,
Breakdown, and Asymptotics of Elastic Theory.}
\author{Alex Buchel, James P. Sethna}
\address{Laboratory of Atomic and Solid State Physics,\\
Cornell University, Ithaca, NY, 14853-2501}
\maketitle

\begin{abstract}

We study a class of models for brittle fracture: elastic theory models
which allow for cracks but not for plastic flow. We show that these models
exhibit, at all finite temperatures, a transition to fracture under applied
load similar to that at a first order liquid-gas transition.  We study this
transition at low temperature for small tension. We discuss the appropriate
thermodynamic limit of these theories: a large class of boundary conditions
is identified for which the energy release for a crack becomes independent
of the macroscopic shape of the material. Using the complex variable method
in a two-dimensional elastic theory, we prove that the energy release in an
isotropically stretched material due to the creation of an arbitrary curvy
cut is the same to cubic order as the energy release for the straight cut
with the same end points.  We find the normal modes and the energy spectrum
for crack shape fluctuations and for crack surface phonons, under a uniform
isotropic tension.  For small uniform isotropic tension in two dimensions we
calculate the essential singularity associated with fracturing the material
in a saddle point approximation including quadratic fluctuations.  This
singularity determines the lifetime of the material (half-life for
fracture), and also determines the asymptotic divergence of the high-order
corrections to the zero temperature elastic coefficients.  We calculate the
asymptotic ratio of the high-order elastic coefficients of the inverse bulk
modulus and argue that the result is unchanged by nonlinearities --- the
ratio of the high-order nonlinear terms are determined solely by the linear
theory.  
\end{abstract}

\pacs{PACS numbers: 03.40.Dz, 46.30.Nz, 62.20.Dc, 62.20.Mk, 64.60.Qb, 
82.60.Nh}

]

\narrowtext

\section{Introduction}

Early in the theory of fracture, Griffith\cite{g} used Inglis' stress
analysis\cite{inglis} of an elliptical flaw in a linear elastic material to
predict the critical stress under which a crack irreversibly grows, causing
the material to fracture.  Conversely, for a stressed solid the Griffith
criterion determines the crack nucleation barrier: if the material has
micro-cracks due to disorder or (less commonly) thermal fluctuations, how
long does a micro-crack have to be to cause failure under a given load?  In
a sense, a solid under stretching is similar to a supercooled gas: the point
of zero external stress plays the role of the liquid-gas condensation point.
Fisher's\cite{fisher} theory of the condensation point predicts that the
free energy of the system develops an essential singularity at the
transition point.  In this paper we develop a framework for the
field-theoretical calculations of the thermodynamics of linear elastic
theory with cracks (voids) that naturally incorporates the quadratic
fluctuations, and we calculate the analogue of Fisher's essential
singularity.  As it is well known, the imaginary part of this essential
singularity can be used to give the lifetime to fracture: what is the rate
per unit volume of a micro-crack fluctuations large enough to nucleate
failure? 

There is much work on thermal fluctuations leading to failure at rather high
tensions, near the threshold for instability (the spinodal
point)\cite{selinger}; there is also work on the role of disorder in
nucleating cracks at low tensions\cite{k1}.  We are primary interested in
the thermal statistical mechanics of cracks under {\it small} tension. We
must admit and emphasize that, practically speaking, there are no thermal
crack fluctuations under small tension --- our calculations are of no
practical significance.  Why are we studying thermal cracks in this formal
limit?  First, for sufficiently small tension, the bulk of the material
(excluding regions near the crack tips) obeys linear elastic theory, thus
making analytical analysis of the fracture thermodynamics tractable.
Second, the {\it real} part of our essential singularity implies that
nonlinear elastic theory is not convergent! Just as in quantum
electrodynamics\cite{dyson} and other field theories\cite{zj}, for all
finite temperatures, nonlinear elastic theory is an asymptotic expansion,
with zero radius of convergence at zero pressure.  We will calculate the
high-order terms in the perturbation expansion governing the response of a
system to infinitesimal tension. We find it intriguing that Hook's law is
actually a first term in the asymptotic series. 

The paper is organized as follows. In the next section, following methods
known in the crack community\cite{b}---\cite{sl}, we carefully examine the
thermodynamic limit of an equilibrium linear elastic theory with voids. We
consider a crack as a special case of a void. We specify the class of
boundary conditions which insure that the energy release is independent of
the shape of the material boundary at infinity and independent of the
prescribed boundary conditions. This is extremely important for the
investigation of the singular structure of the free energy, for the latter
can develop singularities only in the thermodynamic limit \cite{g1}.  Using
the complex variable method in a two-dimensional elastic theory, we
calculate the energy release of an arbitrary curvy crack to quadratic order
in kink angles in section \uppercase\expandafter{\romannumeral3}.  In
section \uppercase\expandafter{\romannumeral4} we find the spectrum and the
normal modes of the boundary fluctuations (surface phonons) of a straight
cut under uniform isotropic tension at infinity. Section
\uppercase\expandafter{\romannumeral5} is devoted to the calculation of the
imaginary part of the free energy.  The calculation of the contribution of
thermal fluctuations depends on the ``molecular structure'' of our material
at short length scales --- in field theory language, it is {\it
regularization} {\it dependent}. We calculate the imaginary part of the free
energy both for $\zeta$-function and a particular lattice regularization,
and determine the temperature dependent renormalization of the surface
tension.  Earlier we showed\cite{we} that the thermal instability of an
elastic material with respect to fracture results in non-analytical behavior
of the elastic constants (e.g. the bulk modulus) at zero applied stress. In
section \uppercase\expandafter{\romannumeral6} we extend the
calculation\cite{we} of the high order expansion of the inverse bulk modulus
by including quadratic fluctuations.  We show there that the asymptotic
ratio of the high order elastic coefficients, written in terms of the
renormalized surface tension, is {\it independent} of regularization (for
the cases we have studied), and we argue also that they are independent of
nonlinear effects near the crack tips. (The asymptotic {\it nonlinear}
coefficients depend only on the {\it linear} elastic moduli.)  In section
\uppercase\expandafter{\romannumeral7} we perform the simplified calculation
(without fluctuations) in several more general contexts: anisotropic strain
(nonlinear Young's modulus), cluster nucleation and dislocation nucleation,
and three-dimensional brittle fracture. We also discuss the effects of vapor
pressure --- nonperturbative effects when bits detach from the crack!
Finally, we summarize our results in section 
\uppercase\expandafter{\romannumeral8}. 

\section{The thermodynamic limit of the energy release}
Elastic materials under a stretching load can relieve deformation energy
through the formation of cracks and voids. The famous Griffith criteria
\cite{g} for a crack propagation is based on the balance between the energy
release and the increase in the material surface energy due to extending the
crack. For a finite size system the energy release is a well defined
quantity that depends on the shape of the material boundary. The situation
becomes more subtle in case of an infinite elastic media. In principle one
can calculate the energy release analyzing stress fields near the crack tips
and thus avoid the necessity of worrying about infinite-sized media. This
method, developed by Irvin in the 1950s, is known as the stress intensity
approach\cite{ewalds}.  Despite its enormous practical importance in
numerical calculations, it is usually of little help in analytical
calculations. To apply the stress intensity factor approach, one has to be
able to compute the stresses near the crack tips, which is possible only in
several simple cases. (The extension of the Irvin's method though the use of
the path independent J-integral\cite{rice}, for example, is applicable only
for cracks with flat surfaces.)  Alternatively, the energy release can be
calculated considering the system as a whole. In this approach, to compute
the energy release one has to evaluate the work done by external forces and
the change in the energy of elastic deformation. The change in the energy of
the elastic deformation involves the difference between two infinitely large
quantities for an infinite material; the latter thus requires some sort of
infinite-volume limit. In this section we discuss the energy release due to
the relaxation of the boundaries of a finite number of voids (cracks) in the
limit of an infinitely large stressed material. The methods of this section
will be used again later in the paper.  

We focus on the energy release calculation for a void formed in an infinite
two-dimensional elastic material. The result is then extended to the case of
a finite number of voids (remember that a crack can be considered as a
degenerate void) and for the energy of void formation in a three dimensional
elastic material.  The energy release for a finite size system with a crack
has been calculated by Bueckner\cite{b} and later generalized by
Rice\cite{rice2} for void formation. Their analysis allow for a large class
of boundary conditions, mixing regions of fixed displacements and fixed
stresses at the perimeter. In what follows we will use a slight modification
of Bueckner's argument.
 
\begin{figure}
\centerline{
\psfig{figure=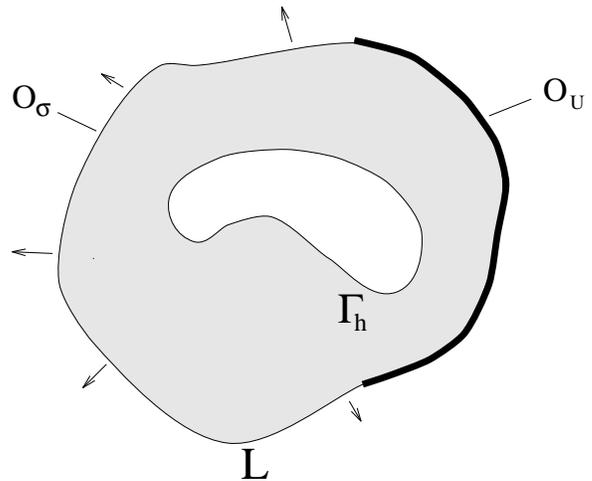,width=3truein}}
{\caption{To calculate the energy release in the thermodynamic 
limit we specify displacements along $O_U$ and stresses along $O_{\sigma}$.}
\label{fr}}
\end{figure}

We define the energy release due to the formation of a void in an infinite
material as the energy in a previously stretched material, released by
cutting out a hole in it and letting the hole boundary relax the stress: the
deformation energy of the discarded piece is excluded. Let's consider an
infinite linear elastic material subject to stress fields
$\sigma_{xx}^{\infty}$, $\sigma_{yy}^{\infty}$ and $\sigma_{xy}^{\infty}$ at
infinity.  We want to calculate the energy release $E_{\rm release}$ due to
cutting out a hole with boundary $\Gamma_h$. We assume that the hole
boundary is stress free and non self-intersecting as a result of the stress
relaxation.  $\Gamma_b$ denotes a regularization boundary with a
characteristic size $L$.  Before the void formation, displacements along
$\Gamma_b$ are $\vec{U}_1^b=(u_1^b,v_1^b)$, while those along the
prospective void contour $\Gamma_h$ correspondingly
$\vec{U}_1^h=(u_1^h,v_1^h)$.  Let $O_{\sigma}$ and $O_U$ be nonintersecting
parts of $\Gamma_b$, such that $O_{\sigma}\cup O_U=\Gamma_b$. Along
$O_{\sigma}$ we fix the stresses $\sigma_{xx}^{\infty}$,
$\sigma_{yy}^{\infty}$ and $\sigma_{xy}^{\infty}$ as a function of position;
along $O_U$ we fix the displacements to be $\vec{U}_1^b$ (Figure \ref{fr}).
In analogy to the usage in field theory, each choice of boundary conditions
$O_\sigma$, $\sigma^{\infty}$, $O_U$, $U^b_1$ we call a {\it regularization}.
For a fixed $L$ we calculate the energy release $E_{\rm release}^L$.  The
thermodynamic limit of the energy release is then given by 
\begin{eqnarray}
E_{\rm release}=\lim_{L\to\infty} E_{\rm release}^L .    
\label{3}
\end{eqnarray} 
Physically the described regularization means that we compute
first the energy release due to the formation of the void $\Gamma_h$ in a
finite size material with the boundary $\Gamma_b$, and then push the outer
(regularization) boundary to infinity.  We will show that defined as above,
$E_{\rm release}$ is independent of a particular choice of $O_{\sigma}$ and
$O_U$, even if they are themselves functions of $L$. The boundary condition
with $O_{\sigma}=\Gamma_b$ is known as a ``fixed tension boundary
condition'', while that with $O_U=\Gamma_b$ is referred to as a ``fixed grip
boundary condition''. 

Let's first consider a straight cut. As shown by explicit
calculation\cite{good}, the ``fixed tension'' and the ``fixed grip''
boundary conditions for the special case of a straight cut of length $\ell$
opened by a uniform isotropic tension $T$, give the same energy release
\begin{equation}
E_{\rm release}={{\pi T^2 {\ell}^2 (\chi+1)}\over{32\mu}}
={{\pi T^2 {\ell}^2}\over {4 Y}} .
\label{st}
\end{equation}
The material elastic constants $\mu$ and $\chi$ can be expressed
through its Young's modulus $Y$ and Poisson ratio $\sigma$ as follows
\begin{eqnarray}
\mu&=&{Y\over {2 (1+\sigma)}} \label{26}\\ 
\chi&=&{{3-\sigma} \over {1+ \sigma} } .\nonumber
\end{eqnarray}
(The given value for $\chi$ corresponds to a plain stress in a three
dimensional elastic theory; for a plain strain one should use
 $\chi=3-4\sigma$.)  Note that (\ref{st}) coincides with Griffith's result.
This result is not trivial! If one calculates the energy change in a region
$\Gamma_b$ embedded in an infinite medium (i.e., neither fixing the
displacements nor allowing them to relax at fixed stress), the energy
release does depend on the shape of the boundary. The relaxations at the
boundary scale like $1/L$: even as $L\to\infty$, integrated over the
perimeter they must be included for a sensible thermodynamic limit.

For a general void, the energy release  is a sum of the work performed by 
the external $(W_e)$ and the internal $(W_i)$ forces as a result of the
void formation 
\begin{eqnarray}
E_{\rm release}^L=W_e+W_i .
\label{4}
\end{eqnarray}
Let $e_{ij}^{(1)}$, $\sigma_{ij}^{(1)}$ be the strain and
stress fields of the first state before the cut is made and $e_{ij}^{(2)}$,
$\sigma_{ij}^{(2)}$ define the fields of the second state with the void;
finally displacements along $O_{\sigma}$ for the second state are
$\vec{U}_2^b=(u_2^b,v_2^b)$ and displacements along the void boundary
$\Gamma_h$ are given by $\vec{U}_2^h=(u_2^h,v_2^h)$.  As the void 
boundary relaxes, the external forces do work 
\begin{eqnarray}
W_e=\int_{O_{\sigma}} \vec{F}_n (\vec{U}_2^b-\vec{U}_1^b) d{\ell}_1  
\label{5}
\end{eqnarray}
where $\vec{F}_n=(F_{n_x},F_{n_y})$ is the traction along $O_{\sigma}$ 
defined through the asymptotic stress fields
\begin{eqnarray}
F_{n_x}=\sigma_{xx}^{\infty} n_{x} +\sigma_{xy}^{\infty} n_{y} 
\label{6} \\
F_{n_y}=\sigma_{xy}^{\infty} n_{x} +\sigma_{yy}^{\infty} n_{y} \nonumber
\end{eqnarray}
with $(n_{x},n_{y})$ being the outwards normal to $O_{\sigma}$.
The work done by  the internal forces is the change in the energy of
the elastic deformation of the first and the second states
\begin{eqnarray}
W_i={1\over2}{\int\int}_A \sigma_{ij}^{(1)} e_{ij}^{(1)} dA -
{1\over2}{\int\int}_A \sigma_{ij}^{(2)} e_{ij}^{(2)} dA .
\label{7}
\end{eqnarray}
The integration in (\ref{7}) is performed over the area $A$ of the 
material excluding the void; the summation over repeating indexes 
is assumed. As a consequence of Hook's law, 
\begin{eqnarray}
{\int\int}_A \sigma_{ij}^{(1)} e_{ij}^{(2)} dA=
{\int\int}_A \sigma_{ij}^{(2)} e_{ij}^{(1)} dA 
\label{8}
\end{eqnarray}
so we can  rewrite (\ref{7}) as 
\begin{eqnarray}
W_i={1\over2}{\int\int}_A (\sigma_{ij}^{(1)} +\sigma_{ij}^{(2)})
(e_{ij}^{(1)}-e_{ij}^{(2)}) dA .
\label{9}
\end{eqnarray}
Let's introduce a plus state specified by the stresses 
 $\sigma^{+}_{ij}=\sigma_{ij}^{(1)} +\sigma_{ij}^{(2)}$ and a minus state 
defined by the strain fields $e^{-}_{ij}=e_{ij}^{(1)}-e_{ij}^{(2)}$. 
Then  the work of the internal forces $(\ref{9})$ is  a mixed energy of
the  plus and the minus states. According to Betti's theorem \cite{betti},
the mixed energy equals  one half the work done  by the stresses
of one state over the displacements of the  other, no matter from 
what state the stresses or displacements are taken. With the stresses
from the plus state and the displacements from the minus state we obtain 
\begin{eqnarray}
W_i&=&{1\over2}\int_{O_{\sigma}} \vec{F}^{+}_n\vec{U}^{-}_b  d{\ell}_1 +
{1\over2}\oint_{\Gamma_h} \vec{F}^{+}_h\vec{U}^{-}_h  d{\ell}_h  
\label{10}\\
\nonumber
&=&{1\over2}\int_{O_{\sigma}} (\vec{F}_n+\vec{F}_n)
(\vec{U}_1^b-\vec{U}_2^b) d{\ell}_1\nonumber\\
&&+{1\over2}\oint_{\Gamma_h} (\vec{F}_h+\vec{0})(\vec{U}_1^h-\vec{U}_2^h) 
d{\ell}_h \nonumber\\
&=&\int_{O_{\sigma}} \vec{F}_n(\vec{U}_1^b-\vec{U}_2^b) d{\ell}_1 +
{1\over2}\oint_{\Gamma_h}\vec{F}_h(\vec{U}_1^h-\vec{U}_2^h) d{\ell}_h
\nonumber
\end{eqnarray}  
where a  traction of the first state $\vec{F}_h$ is
defined as in (\ref{6}) through the stresses of the first state 
along $\Gamma_h$. From (\ref{4}), (\ref{5}) and (\ref{10}) we find 
the regularized energy release\cite{b},\cite{rice2}
\begin{eqnarray}
E_{\rm release}^L={1\over2}\oint_{\Gamma_h}\vec{F}_h(\vec{U}_1^h-
\vec{U}_2^h) d{\ell}_h\label{11} .
\end{eqnarray}
The thermodynamic limit (\ref{3}) is 
then taken by  simply replacing  $\vec{U}_2^h$ with its value
\begin{eqnarray}
\vec{U}_{\infty}^h=\lim_{L\to\infty}\vec{U}_2^h
\label{12}
\end{eqnarray}
for the infinite media. One can check (using for example the 
complex variable method described below) that the difference between 
two can at most be  of  order $O(1/L)$, thus according to (\ref{11}) 
and (\ref{3}) not contributing to $E_{\rm release}$. So we conclude that
\begin{eqnarray}
E_{\rm release}={1\over2}\oint_{\Gamma_h}\vec{F}_h(\vec{U}_1^h-
\vec{U}_{\infty}^h) d{\ell}_h .
\label{13}
\end{eqnarray}
Thus the {\it total} energy release is half the work 
done at the cut boundary.
The above result is explicitly independent of a particular 
regularization from the discussed class as well as of the shape of
the regularization contour $\Gamma_b$. 
It  has a straightforward extension for a finite number of voids
in a media. For a media with $n$ holes with contours 
 $\lbrace \Gamma_{h_i} \rbrace$, (\ref{13}) generalizes to 
\begin{eqnarray}
E_{\rm release}={1\over2}\sum_{i=1}^n \oint_{\Gamma_{h_i}}\vec{F}_{h_i}
(\vec{U}_1^{h_i}-\vec{U}_{\infty}^{h_i})d{\ell}_{h_i} .
\label{18}
\end{eqnarray} 
The same class of regularizations gives well defined thermodynamic limit in
a three dimensional case as well.  The arguments are practically the same
resulting in the 
 energy release (\ref{13})  with the only change in the use of 
surface integrals  rather than the contour ones for the regularization   
and  void boundaries. 

In order to use (\ref{13}) one has to know displacements along the void
contour for the second state. For all but several simple cases this is a
very complicated problem to approach analytically: at best, one can hope to
get an asymptotic behavior of the displacement and stress fields.  So for
practical purposes we have to find an asymptotic analog of the above
expression that would allow the calculation of the energy release from the
asymptotic behavior of the displacement fields.  To do this, it is
convenient to specify the stresses at infinity --- the regularization with
$O_U=\emptyset$ and $O_{\sigma}=\Gamma_b$. As we have already shown, this
does not restrict the applicability of the result.  To get the asymptotic
expression for the energy release we first note that elastic states one and
two are each in equilibrium, so from Clapeyron's theorem\cite{betti}
\begin{eqnarray}
{1\over2}{\int\int}_A \sigma_{ij}^{(1)}e_{ij}^{(1)} dA &=& 
{1\over2}\oint_{\Gamma_b}
\vec{F}_n\vec{U}_1^b d{\ell}_b +{1\over2}\oint_{\Gamma_h}\vec{F}_h
\vec{U}_1^h d{\ell}_h 
\nonumber\\
\label{14}
{1\over2}{\int\int}_A \sigma_{ij}^{(2)}e_{ij}^{(2)} dA &=& 
{1\over2}\oint_{\Gamma_b}\vec{F}_n\vec{U}_2^b d{\ell}_b .
\end{eqnarray}    
Which, following (\ref{4}), (\ref{5}) and (\ref{7}) gives
\begin{eqnarray}
E_{\rm release}^L&=&{1\over2}\oint_{\Gamma_b}\vec{F}_n\vec{U}_2^b 
d{\ell}_b -{1\over2}\oint_{\Gamma_b}\vec{F}_n\vec{U}_1^b 
d{\ell}_b\nonumber\\ 
&&+{1\over2}\oint_{\Gamma_h}\vec{F}_h\vec{U}_1^h d{\ell}_h .
\label{15}
\end{eqnarray}  
Next, we break $\vec{U}_2^b$ into two pieces:  displacements of 
the second elastic state for the infinite media along $\Gamma_b$,
$\vec{U}_f^b$, and   the {\it boundary relaxation} displacements,
$\vec{U}_r^b$,
\begin{equation}
\vec{U}_2^b=\vec{U}_f^b+\vec{U}_r^b.
\label{15f}
\end{equation}
The {\it boundary relaxation} displacements comes from relaxing the 
the stresses along $\Gamma_b$  for the infinite material to 
comply with the thermodynamic limit prescription: the ``fixed 
tension boundary condition''. For  the energy release calculation 
it is sufficient to relax the stresses to $O(1/L^3)$ --- the
stresses of order $O(1/L^3)$ generate the relaxation displacements 
along $\Gamma_b$ of order $O(1/L^2)$ and thus do not contribute to the 
energy release in the limit $L\to\infty$.
So, from (\ref{3}), (\ref{15}) and (\ref{15f})
\begin{eqnarray}
E_{\rm release}=\lim_{L\to\infty}\biggl\lbrace
&&{1\over2}\oint_{\Gamma_b}\vec{F}_n(\vec{U}_f^b
+\vec{U}_r^b) d{\ell}_b 
-{1\over2}\oint_{\Gamma_b}\vec{F}_n\vec{U}_1^b d{\ell}_b\nonumber\\
&&+{1\over2}\oint_{\Gamma_h}\vec{F}_h\vec{U}_1^h d{\ell}_h
\biggr\rbrace .
\label{17}
\end{eqnarray}
This is the desired expression. One can see that (\ref{17}) is nothing but
the difference between the regularized elastic deformation energy of the
infinite media with the void (the first integral) and the elastic energy
stored in the same media before the formation of void, not accounting for
the elastic energy of the void itself (the remaining two integrals).
Although the asymptotic expression for the energy release may look more
complicated than (\ref{13}) and actually requires solving of two elastic
problems, its use is justified because it is difficult to find displacements
of the second state along the void contour.

\section{Energy release of ``slightly'' curvy cuts}
Armed with the knowledge that the energy release is independent of the shape
of the regularization boundary, we now turn to the calculation of the energy
release due to the opening of a ``slightly'' curvy cut $\Gamma_h$ in a two
dimensional isotropic linear elastic infinite media subject to a uniform
isotropic tension $T$ at infinity.  We find an amazingly simple answer:  the
energy release of a curvy cut coincides to cubic order with the energy
release of the straight cut with the same end points.

An elastic state is completely defined
once displacements $(u,v)$ are known everywhere. Rather than
considering  these two functions,  Muskhelishvili\cite{m}  introduces two 
complex functions $\phi(z)$ and $\psi(z)$ that in equilibrium 
 should be the functions of only one complex
variable $z$ (i.e. do not depend  $\overline{z}$). 
Moreover in our case (a uniform isotropic tension at infinity)
$\phi(z)$  decomposes as
\begin{eqnarray}
\phi(z)&=&{1\over2}T z + \phi_0(z) \label{35}
\end{eqnarray} 
The functions $\phi_0(z)$ and $\psi(z)$ are  holomorphic
in the complex $z$ plane including infinity but excluding the cut 
contour.  This description associates the components 
of stress $( \sigma_{xx},\sigma_{yy},\sigma_{xy})$ and 
displacement $(u,v)$ to $(\phi,\psi)$ by the following relations
\begin{eqnarray}
\sigma_{xx}+\sigma_{yy}&=&2 (\phi'(z)+\overline{\phi'(z)})\label{36}\\
\sigma_{yy}-\sigma_{xx}+2i\sigma_{xy}&=&2(\overline{z}\phi''(z)+\psi'(x))
\nonumber\\
2\mu(u+iv)&=&\chi\phi(z)-z\overline{\phi'(z)}-\overline{\psi(z)}
\nonumber
\end{eqnarray}
(The detailed discussion of the change of ``variables'' $(u,v)\to(\phi,\psi)$
along with the derivation of (\ref{35})-(\ref{36}) can be found in \cite{m}.)
Using a circular regularization contour ($\Gamma_b$ is a circle of radius
$L$) and an expression analogous to (\ref{17}), Sih and Liebowitz\cite{sl}
explicitly computed stresses along the outer boundary $\Gamma_b$ and found
the energy release
\begin{eqnarray}
E_{\rm release}=-{{\pi T}\over{4\mu}}(1+\chi)\rm{Re}[y_1] 
\label{60}
\end{eqnarray}
with $y_1$ being the residue of $\psi(z)$ at infinity (the $1/z$ coefficient
in the expansion about $z=\infty$).
Of course, to determine $y_1$ we still need to solve the asymptotics of
the elasticity problem.

\begin{figure}
\centerline{
\psfig{figure=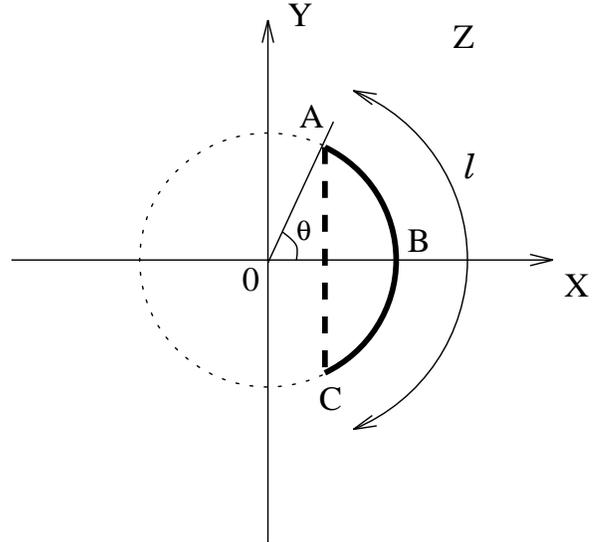,width=3truein}}
{\caption{The ``smile''-like   cut  ABC and the straight 
cut AC result in the same energy release to cubic order
 in $\theta$.}
\label{f1}}
\end{figure}

To illustrate the correspondence between  the energy release of a curvy cut 
and the straight one with the same end points, let's consider a rare 
example where it is possible to find an  exact analytical solution. 
Suppose a material with a 
``smile'' cut  - an arc of a circle ABC - of  total arc length 
$\ell$ (Figure \ref{f1})
is subject to a uniform isotropic stretching $T$ at infinity. Expanding 
the exact answer in\cite{m} about $z=\infty$, we find 
\begin{equation}
\psi_{\rm ABC}(z)=-{{T \ell^2}\over {8
z}}{{8\sin^2\theta/2}
\over{\theta^2(3-\cos\theta)}}+O\biggl({{1}\over{z^2}}\biggr)
\label{61}
\end{equation} 
which according to (\ref{60}) gives the energy release $E_{\rm ABC}$
\begin{equation}
E_{\rm ABC}={{\pi T^2 \ell^2}\over {32
\mu}}(1+\chi){{8\sin^2\theta/2}
\over{\theta^2(3-\cos\theta)}} .
\label{62}
\end{equation}   
On the other hand, for a straight cut AC of length 
 $(\ell/\theta)\sin\theta$, the holomorphic function $\psi_{\rm AC}(z)$
has asymptotic behavior \cite{m}
\begin{equation}
\psi_{\rm AC}(z)=-{{T \ell^2}\over {8  z}}{{\sin^2\theta}
\over{\theta^2}}+O\biggl({{1}\over{z^2}}\biggr)
\label{63}
\end{equation} 
resulting in the energy release $E_{\rm AC}$ 
\begin{equation}
E_{\rm AC}={{\pi T^2 \ell^2}\over {32
\mu}}(1+\chi){{\sin^2\theta}\over{\theta^2}} .
\label{64}
\end{equation} 
For small $\theta$ we find from (\ref{62}) and (\ref {64}) the advertised 
result: the energy release of ABC coincides with that of AC 
to cubic order in $\theta$, but not to quartic order!
\begin{eqnarray}
E_{\rm ABC}&=&{{\pi T^2 \ell^2}\over {32
\mu}}(1+\chi)\biggl(1-{{\theta^2}\over3}+{{77\theta^4}\over{720}}
+O(\theta^6)\biggr)
\label{65}\\
E_{\rm AC}&=&{{\pi T^2 \ell^2}\over {32
\mu}}(1+\chi)\biggl(1-{{\theta^2}\over3}+{{2\theta^4}\over{45}}
+O(\theta^6)\biggr) . \nonumber
\end{eqnarray}   

We proceed now with the general proof. First, an arbitrary cut is
approximated by a finite number of line segments, parameterized by the kink
angles $\alpha_i$ --- the angles between consecutive kinks. The exact shape
of the cut is then restored as the length of each link goes to zero (as
their number goes to infinity). The energy release is evaluated to cubic
order in the kink angles, e.g for the $n$ - kink regularization, the energy
release $E_n(\lbrace\alpha_i\rbrace)$ for a curvy cut with a fixed
separation $\ell_p$ between the endpoints is approximated as 
\begin{eqnarray}
E_n(\lbrace\alpha_i\rbrace) &=& E^{(0)} + \sum_{i=1}^n E_i^{(1)}\alpha_i
+ \sum_{i=1}^n\sum_{j=1}^n E^{(2)}_{ij} \alpha_i\alpha_j\nonumber\\ 
&&+ \sum_{i=1}^n\sum_{j=1}^n \sum_{m=1}^n E^{(3)}_{ijm} \alpha_i\alpha_j
\alpha_m+O(\alpha_i^4)
\label{66}
\end{eqnarray} 
where $E^{0}$ is the energy release for a straight cut of
length $\ell_p$ and the coefficients $E_i^{(1)}$, $E^{(2)}_{ij}$ and
$E^{(3)}_{ijm}$ depend only on the positions of the kinks along the cut.  We
claim that all coefficients up to cubic order are zero, and thus the energy
of a curvy cut and the straight one with the same endpoints can differ only
at $O(\alpha_{i}^4)$.

That $E_i^{(1)}$ and $E_{ijm}^{(3)}$ (in fact, all terms odd in the kink
angles) are zero follows from a symmetry argument: cuts (having the same
number of segments with the corresponding segments being of the same length)
with kink angles $\lbrace\alpha_i\rbrace$ and $\lbrace -\alpha_i\rbrace$
respectively, are mirror images to each other with respect to the first
link. The boundary condition for our problem (a uniform tension at infinity)
is reflection invariant, so 
\begin{eqnarray}
E_n(\lbrace\alpha_i\rbrace)=E_n(\lbrace -\alpha_i\rbrace)
\label{67}
\end{eqnarray}
which requires that all energy release terms odd in the kink angles vanish.
To calculate $E_{ij}^{(2)}$ for a given pair of indexes we can put all kink
angles to zero except for $\alpha_{i}$ and $\alpha_{j}$, reducing the
$n$-kink problem to a two-kink one.  From now on we will consider only the
two-kink problem to quadratic order in the kink angles. 

\begin{figure}
\centerline{
\psfig{figure=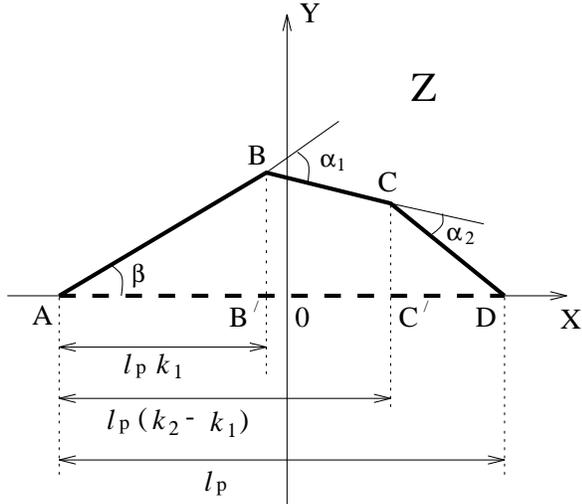,width=3truein}}
{\caption{The two-kink cut  ABCD can be considered as a
 {\it deformation} of a straight cut AD.}
\label{f2}}
\end{figure}

 We choose the coordinate system $XY$ in the complex 
 $z$ plane in such a way that the ends of the two-kink cut are on the
 $X$ axis, symmetric with respect to the $Y$ axis (Figure \ref{f2}).  
Assuming a uniform isotropic tension  $T$ at infinity we rewrite
 (\ref{60}), explicitly indicating the dependence of the energy release on 
the kink angles 
\begin{eqnarray}
E_2(\alpha_1,\alpha_2)=-{{\pi T}\over{4\mu}}(1+\chi)\rm{Re}
[y_1(\alpha_1,\alpha_2)]
\label{68}
\end{eqnarray}
where $y_1(\alpha_1,\alpha_2)$ is $1/z$ coefficient in the expansion of the
function $\psi(z)$ at infinity. As discussed earlier in the section,
$\psi(z)$ is a holomorphic function in the complex $z$ plane including
infinity (the extended complex plane) but excluding the two-kink cut.  The
other function $\phi(z)$ that is necessary for the specification of the
equilibrium elastic state satisfies (\ref{35}) with $\phi_0(z)$ holomorphic
in the same region as $\psi(z)$. The analytical functions $\phi(z)$ and
$\psi(z)$ must provide a stress free cut boundary, which following\cite{m}
can be expressed as
\begin{equation}
i f(\phi,\psi)=\bigg[\phi(z)+z\overline{\phi'(z)}+\overline{\psi(z)}
\bigg]\Bigg|_W^X=0
\label{70}
\end{equation}
where $f=F_x+iF_y$ is the complex analog of the force acting on the 
portion of the cut boundary  between points $W$ and $X$.  

It is important to note that any two pairs of functions 
$(\phi_0^1(z),\psi^1(z))$ and $(\phi_0^2(z),\psi^2(z))$ 
that are holomorphic in the 
extended $z$ plane  excluding the same curvy cut, 
and which provide the stress free cut boundaries  to 
$O(\alpha^3)$ (\ref{70}), 
can differ only by $O(\alpha^3)$ everywhere:
\begin{eqnarray}
\delta\phi(z)&=&\phi_0^1(z)-\phi_0^2(z)=O(\alpha^3)\label{71}\\
\delta\psi(z)&=&\psi^1(z)-\psi^2(z)=O(\alpha^3) .\nonumber
\end{eqnarray}
This follows explicitly from Cauchy's theorem, but also follows from 
the elastic theory. 
Each pair $(\phi_0^1(z)+T z/2,\psi^1(z))$ or 
$(\phi_0^2(z)+T z/2, \psi^2(z))$ defines the equilibrium elastic state 
with  stresses of order $O(\alpha^3)$
along the cut boundary and  uniform isotropic stretching $T$ at 
infinity. So, $(\delta\phi(z),\delta\psi(z))$ corresponds to the 
equilibrium  state with the specified stresses of order 
$O(\alpha^3)$ along the cut boundary and zero tension at infinity.
Thus (\ref{71}) follows because the response to this force within 
linear elastic theory must be linear.
The above argument guarantees that once we find  
$\phi_0(z)$ and $\psi(z)$ that satisfy 
the discussed constraints to $O(\alpha^3)$, we can use them to calculate  
the energy release of the curvy cut to quadratic order. 
  
\begin{figure}
\centerline{
\psfig{figure=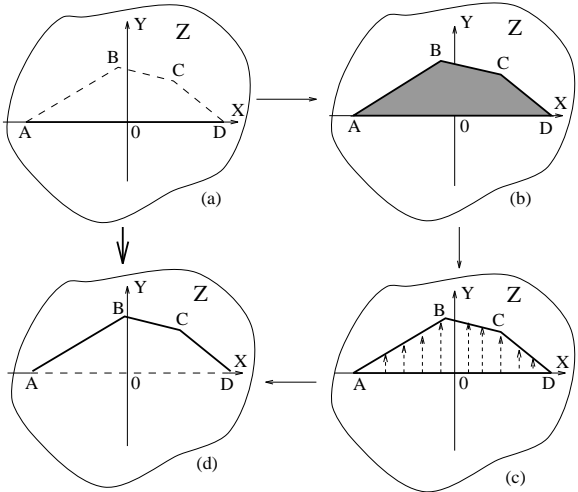,width=3truein}}
{\caption{Functions $\phi_0^s(z)$ and $\psi^s(z)$ holomorphic 
in the complex $z$ plane excluding the straight cut AD (a), 
can be ``made'' holomorphic in the complex $z$ plane 
excluding the two-kink cut ABCD (d).}
\label{f3}}
\end{figure}

Let   functions $\phi^s(z)$ and  $\psi^s(z)$  define the 
equilibrium elastic state of a material with a  straight cut AD 
subject to a uniform tension $T$ at infinity. $\phi_0^s(z)=\phi^s(z)-T z/2$ 
and  $\psi^s(z)$ should then be holomorphic in the extended 
complex $z$ plane excluding the straight  cut and  should provide 
stress free boundaries along AD. 
Muskhelishvili finds\cite{m} 
\begin{eqnarray}
{{d\phi^s(z)}\over{dz}}&=&{T\over2} {z\over{\sqrt{z^2-{\ell_p}^2/4}}}
\label{72}\\
{{d\psi^s(z)}\over{dz}}&=&{T\over8}{{z{\ell_p}^2}\over{
(z^2-{\ell_p}^2/4)\sqrt{z^2-{\ell_p}^2/4}}} .
\nonumber
\end{eqnarray}
(To obtain $\phi^s(z)$ and $\psi^s(z)$ we integrate (\ref{72}); the
arbitrariness in the integration constants reflect the ambiguity in the
displacements up to a rigid motion of the material as a whole.)  Note that
$\phi_0^s(z)$ and $\psi^s(z)$ can be ``made'' holomorphic everywhere in the
complex $z$ plane excluding the two-kink cut ABCD, and thus can serve as a
good starting point for the construction of $\phi_0(z)$ and $\psi(z)$.  The
process of an analytical continuation is demonstrated by Figure \ref{f3}.
$\phi_0^s(z)$ (or equivalently $\psi^s(z)$) is holomorphic in the $z$ plane
excluding the straight cut AD (4a). Removing the region ABCDA (4b), we make
it holomorphic in the complex plane excluding ABCDA. Now we analytically
continue $\phi_0^s(z)$ from the link AD (4c) into the removed region (the
continuation is possible explicitly using (\ref{72})).  The obtained
function becomes holomorphic everywhere in the complex $z$ plane excluding
the two-kink cut ABCD (4d), moreover the original function and the one
obtained through the analytical continuation coincide outside ABCDA. 

\begin{figure}
\centerline{
\psfig{figure=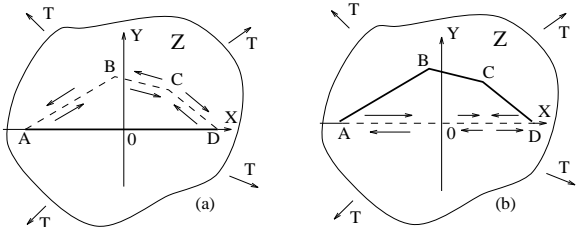,width=3truein}}
{\caption{The stress free boundary of a two--kink
 ABCD cut (b) can be mimicked by applying the tangential 
force to the previously unstressed (a) straight cut boundary AD.}
\label{f5}}
\end{figure}

The idea of construction the holomorphic functions $\phi_0(z)$ and $\psi(z)$
is simple: we start with the functions $\phi_0^s(z)$ and $\psi^s(z)$ and
calculate to quadratic order the stresses along the two-kink cut boundary
ABCD under the analytical continuation as described by Figure \ref{f3}. The
stresses along the curvy cut boundary (Figure \ref{f5}) are then compensated
up to quadratic order in the kink angles by introducing counter-forces along
the original (straight) cut, leading to corrected functions
$\delta\phi^c(z)$ and $\delta\psi^c(z)$, where
$\phi(z)=\phi^s(z)+\delta\phi^c(z)+O(\alpha^3)$ and
$\psi(z)=\psi^s(z)+\delta\psi^c(z)+O(\alpha^3)$.  For the calculation of the
energy release (\ref{68}) we need the real part of the residue of $\psi(z)$
at infinity: we will show that the residue of $\delta\psi^c(z)$ at infinity
is zero and thus the residues of $\psi(z)$ and $\psi^s(z)$ at $z=\infty$ are
the same --- which means that the energy release for the curvy cut ABCD is
the same as that of for the straight cut AD.
    
Let's assume that points $W$ and $X$ are on the upper boundary of the link
AB.  From Figure \ref{f2}, $z=t+i\beta(t+\ell_p/2)+O(\alpha^3)$, where $t\in
{\rm AB'}$ and $\beta=(1-k_1)\alpha_1+(1-k_2)\alpha_2+O(\alpha^3)$.  Using
(\ref{70}) we find 
\begin{eqnarray}
&i& f(\phi_0^s,\psi^s)=\bigg[{\phi^s(t)}^++t\overline{{{\phi^s(t)}'}^+}+
\overline{{\psi^s(t)}^+}\bigg]\Bigg|_W^X
\nonumber\\
&&+i\beta(t+\ell_p/2)\biggl({{\phi^s(t)}'}^++\overline{{{\phi^s(t)}'}^+}
\nonumber\\
&&\qquad -t\overline{{{\phi^s(t)}''}^+}-\overline{{{\psi^s(t)}'}^+}\biggr)
\bigg|_W^X\nonumber
\\&&-{{\beta^2(t+\ell_p/2)^2}\over2}\biggl({{\phi^s(t)}''}^+
+t\overline{{{\phi^s(t)}'''}^+}-2\overline{{{\phi^s(t)}''}^+}
\nonumber\\&&\qquad +\overline{{{\psi^s(t)}''}^+}\biggr)\bigg|_W^X
+O(\alpha^3)\nonumber\\
&=&2\beta^2(t+\ell_p/2)^2 {{\phi^s(t)}''}^+\bigg|_W^X
+O(\alpha^3) ,\label{73}
\end{eqnarray}
where $t$ runs along ${\rm AB'}$; the $+$ superscript 
means that the values of $\phi_0^s(t)$ and $\psi(t)$ should 
be taken at the upper boundary of the straight cut.
To obtain the second expression in (\ref{73})
one can plug in the explicit form (\ref{72}), 
or - more elegantly - note  that for $t\in{\rm AB'}$,
$\phi^s(t)$ is pure imaginary and $\psi^s(z)'=-z\phi^s(z)''$.
Either way, it  follows that  the functions $\delta\phi^c(z)$ 
and $\delta\psi^c(z)$  satisfy
\begin{eqnarray}
i\delta f&=&\bigg[{\delta\phi^c(t)}^++t\overline{{{\delta\phi^c(t)}'}^+}+
\overline{{\delta\psi^c(t)}^+}\bigg]\Bigg|_W^X\nonumber\\
&=&-2\beta^2(t+\ell_p/2)^2 {{\phi^s(t)}''}^+\bigg|_W^X .\label{74} 
\end{eqnarray}
This is the force we need to add along the straight cut just below segment
AB to cancel the stress along the curvy cut.  Similar expressions can be
found for the forces needed below BC and CD.  To find $\delta\phi^c(z)$ and
$\delta\psi^c(z)$ we have to solve the elasticity problem for the material
with the straight cut AD, subject to these applied forces $i\delta f$ along
the cut boundary. Fortunately, this problem allows a closed analytical
solution\cite{m}. Expanding the exact expression for $\delta\psi^c(z)$
in\cite{m} we find 
\begin{equation}
\delta\psi^c(z)={{\ell_p}\over{4 \pi iz}}\oint_{\gamma}
{\rm Re}\biggl[i\delta f(x(\sigma))\biggr]d\sigma
+O\biggl({1\over {z^2}}\biggr)\label{75}
\end{equation} 
where the integration is along the unit circle $\gamma$ in the complex
plane, and $i\delta f$ is a function of a variable point
$x(\sigma)=\ell_p(\sigma+1/\sigma)/4$ along the straight cut boundary AD.
Notice from (\ref{74}) that $i \delta f$ is pure imaginary evaluated on the
upper boundary of the link AB:  in this case $|t|<\ell_p/2$, and so the
argument of the square root in (\ref{72}) is negative resulting in pure
imaginary ${\phi^s(t)}'$, thus ${\phi^s(t)}''$ is also pure imaginary. In
fact, as it can be checked explicitly, ${\rm Re}\biggl[i\delta f\biggr]=0$
for arbitrary $W$ and $X$ along the cut boundary. So we conclude from
(\ref{75}) that the residue of $\delta\psi^c(z)$ at infinity is zero and thus
the energy release for the curvy cut ABCD is the same as for the straight
cut AD. The underling physical reason for this seeming remarkable
coincidence is that to imitate the stress free curvy cut to quadratic order
in the kink angles we have to apply only tangential force along the straight
cut (pure imaginary $i\delta f$ means $F_y=0$), which do no work because a
straight cut under a uniform isotropic tension at infinity opens up but does
not shrink\cite{m}.  

We find that the energy release $E(\ell_p)$ of the curvy cut with 
projected distance 
$\ell_p$ between the endpoints is the same to quadratic order in the kink 
angles as the energy release of the straight cut of length $\ell_p$. 
The latter one is given by the second formula in (\ref{65}) with 
$\theta=0$ (it also coincides with Griffith's result (\ref{st}))
\begin{equation}
E(\ell_p)={{\pi T^2\ell_p^2}\over{32\mu}}(1+\chi) .
\label{c1}
\end{equation}  
The natural variables to describe the curvy cut are its total length $\ell$
and its curvature $k(x),\ x\in [0,\ell]$.  In what follows we express
(\ref{c1}) in these variables and find the normal modes of the curvature
that diagonalize the energy release. 

For the  two-kink cut ABCD (Figure \ref{f2}) of total length $\ell$
one can find
\begin{eqnarray}
\ell_p&=&\ell\biggl(1-{{x_1(\ell-x_1)}\over{2\ell^2}}\alpha_1^2
-{{x_1(\ell-x_2)}\over{\ell^2}}\alpha_1\alpha_2\nonumber\\
&&-{{x_2(\ell-x_2)}\over{2\ell^2}}\alpha_2^2\biggr)+O(\alpha^3)
\label{c2}
\end{eqnarray}
where $x_1$ and $x_2$ parameterize the kink positions:
the length of the link AB is assumed to be $x_1$ and the 
length of the segment ABC  equals $x_2$. 
Similarly, for the n-kink cut of  total length 
$\ell$ with the kink angles $\lbrace\alpha_i\rbrace$ parameterized 
by their distance $\lbrace x_i\rbrace$ from the cut end
\begin{eqnarray}
\ell_p&=&\ell\biggl(1-{1\over 2}\sum_{i,j=1}^n\alpha_i\alpha_j
\biggl[-{{x_ix_j}\over{\ell^2}}+{{{\rm min}(x_i,x_j)}\over\ell}
\biggr]\biggr)\nonumber\\
&&+O(\alpha^3) .\label{c3}
\end{eqnarray}    
Expressing the kink angles through the local curvature of the curve,
$\alpha_i=k(x_i)\Delta x_i/\lambda$, we find the continuous 
limit of (\ref{c3})
\begin{eqnarray}
\ell_p&=&\ell\biggl(1-{1\over 2}\int_{0}^{\ell}\int_{0}^{\ell}
{{dxdy}\over{\lambda^2}} k(x)M(x,y)k(y)\biggr)\nonumber\\
&&+O({k(x)}^3) ,\label{c4}
\end{eqnarray}
with
\begin{equation}
M(x,y)=-{{x y}\over{\ell^2}}+{{{\rm min}(x,y)}\over\ell}
\label{c5}
\end{equation}
and the scale $\lambda$ is introduced to make the curvature dimensionless
($\lambda$ can be associated with the ultraviolet cutoff of the theory ---
roughly the interatomic distance).  Substituting (\ref{c5}) into (\ref{c1})
we find the energy release $E(\ell,k(x))=E(\ell_p)$ of the curvy cut in its
intrinsic variables
\begin{eqnarray}
E(\ell&,&k(x))=\nonumber\\
&&{{\pi T^2\ell^2}\over{32\mu}}(1+\chi)
\biggl(1-\int_{0}^{\ell}\int_{0}^{\ell}{{dxdy}\over{\lambda^2}} 
k(x)M(x,y)k(y)\biggr)\nonumber\\
&&+O({k(x)}^3) .\label{c6}
\end{eqnarray} 
To find the normal modes of the curvature we have to find the eigenvalues
and  eigenvectors of the operator $M(x,y)$. If $k_n(x)$ is 
an eigenvector of $M(x,y)$ with  eigenvalue $\lambda_n$, then
\begin{equation}
\lambda_n k_n(x)=\int_{0}^{\ell}{{dy}\over\lambda}M(x,y)k_n(y) .
\label{c7}
\end{equation}
 From (\ref{c5}), $M(0,y)=0$ and $M(\ell,y)=0$ for arbitrary
$y\in[0,\ell]$, so from (\ref{c7}) the eigenvectors of 
$M(x,y)$ must be zero at $x=0$ and $x=\ell$: $k_n(0)=k_n(\ell)=0$.
An arbitrary function $k_n(x)$ with this property is given by the 
Fourier series
\begin{equation}
k_n(x)=\sum_{m=1}^{\infty}c_m\sqrt{{{2\lambda}\over{\ell}}}\sin
{{\pi m x}\over{\ell}}
\label{c8}
\end{equation}
where the overall constant $\sqrt{2\lambda/\ell}$ is introduced to 
normalize the Fourier modes with the integration measure $dx/\lambda$
over $x\in[0,\ell]$. One can explicitly check from (\ref{c7}) that each 
Fourier mode $\sqrt{2\lambda/\ell}\sin(\pi m x/\ell)$ is in fact an 
eigenvector of  $M(x,y)$ with the eigenvalue $\lambda_m
=\ell/(\pi^2 m^2\lambda)$. In terms of the amplitudes 
of the normal modes $\lbrace c_n\rbrace$, (\ref{c6}) is rewritten as 
\begin{equation}
E(\ell,\lbrace c_n\rbrace)={{\pi T^2\ell^2}\over{32\mu}}(1+\chi)
\biggl(1-\sum_{n=1}^{\infty}{{\ell}\over{\pi^2 n^2\lambda}}
c_n^2\biggr)+O(c_n^3) .\label{c9}
\end{equation}
(\ref{c9}) is the main results of the section: we've calculated  the 
energy release of an arbitrary curvy cut in its intrinsic variables ---
the total length $\ell$ and the curvature $k(x),\ x\in[0,\ell]$ ---
to quadratic order in $k(x)$ and found the normal modes of the curvature 
that diagonalize the energy release. 

In conclusion we mention that the measure in the kink angle space is
Cartesian --- $\prod_i d\alpha_i$ --- (and thus the functional measure $D
k(x)\equiv D \alpha(x)$ is Cartesian), so the measure in the vector space of
the amplitudes of the normal modes $\lbrace c_n\rbrace$ is also Cartesian
--- $\prod_{n=1}^{\infty} dc_n$, because the Fourier transformation $\lbrace
k(x)\rbrace\to\lbrace c_n\rbrace$ is orthonormal. This will be important in
section V, where we will be integrating over crack shapes.

\section{Surface phonons}
In the previous sections we have extensively discussed the calculation of
the energy release due to the equilibrium opening of a cut in an elastic
material. Since our goal is to deal with cracks as thermal fluctuations, we
must also deal with the more traditional elastic fluctuations --- phonons,
or sound. We find here that the bulk fluctuations decouple from the new
surface phonon modes introduced by the cut. We discuss the quadratic
fluctuations for linear elastic material with a straight cut of length
$\ell$ subject to a uniform isotropic tension $T$ at infinity; more
specifically, we calculate the energy release for the material with an
arbitrary opening of the straight cut and we find collective coordinates
(normal modes) that diagonalize the change in the energy. 

An elastic state of the material can be defined through the specification of
its displacements $\vec{U}=(u,v)$ at every point $(x,y)$. For the material
with a cut, the fields $u(x,y)$ and $v(x,y)$ can in principle have a
discontinuity along the cut: assuming that the cut is an interval
 $(x,y)=([-\ell/2,\ell/2],0)$, 
\begin{equation}
2 g_x(x)=u(x,0+)-u(x,0-), \mbox{\hspace{0.1in} }
{\rm for} \mbox{\hspace{0.1in} }x\in[-\ell/2,\ell/2]
\label{78} 
\end{equation}
and
\begin{equation}
2 g_y(x)=v(x,0+)-v(x,0-), \mbox{\hspace{0.1in} }
{\rm for} \mbox{\hspace{0.1in} }x\in[-\ell/2,\ell/2]
\label{79} 
\end{equation}  
may be nonzero. It is clear that the arbitrary state $\vec{U}$ can be
decomposed into the superposition of two states $\vec{U}_g=(u_g,v_g)$ and
$\vec{U}_c=(u_c,v_c)$, where $\vec {U_g}$ is the equilibrium state for given
displacement discontinuity $(g_x,g_y)$ at the cut boundary
(\ref{78}-\ref{79}) and tension $T$ at infinity that maximizes the energy
release, and $\vec{U}_c$ given by $(u_c,v_c)=(u-u_g,v-v_g)$ is a continuous
displacement field everywhere.  Recall that the energy release is the sum of
the work done by the external forces and the work done by the internal
forces (\ref{4}).  We define the energy release $E$ for the elastic state
$\vec{U}$ with respect to the equilibrium state of the material
$\vec{U}_0=(u_0,v_0)$ without the cut under the same loading at infinity, as
a limit of this difference for finite size samples with boundary $\Gamma_b$
and enclosed area $A$. We find following (\ref{4}), (\ref{5}) and (\ref{6})
\begin{equation}
E=\oint_{\Gamma_b} T \vec{n}(\vec{U}-\vec{U}_0) d\ell
+{1\over 2}\int\int_A (\sigma_{ij}^0 e_{ij}^0 -\sigma_{ij} e_{ij}) d A
\label{81}
\end{equation} 
where $\sigma_{ij}^0$ and $e_{ij}^0$ are the stresses and strains of the
equilibrium elastic state of the uncracked material $\vec{U}_0$;
$\sigma_{ij}$ and $e_{ij}$ are the stresses and strains of the elastic state
of material with the straight cut and displacement field $\vec{U}$; and
$\vec{n}$ is a unit normal pointing outwards from the regularization
boundary $\Gamma_b$. This argument is similar to that in the second section,
but the elastic state $\vec{U}$ is not an equilibrium one and so the
arguments there are not directly applicable.  We rewrite the energy release
(\ref{81}) making use of the decomposition $\vec{U}=\vec{U}_g+\vec{U}_c$ to
get 
\begin{eqnarray}
E&=&\oint_{\Gamma_b} T \vec{n}(\vec{U}_g-\vec{U}_0) d\ell
+{1\over 2}\int\int_A (\sigma_{ij}^0 e_{ij}^0 -\sigma_{ij}^g e_{ij}^g) d A
\nonumber\\
&&-{1\over 2}\int\int_A \sigma_{ij}^c e_{ij}^c  d A\nonumber\\
&&+\oint_{\Gamma_b} T \vec{n}\vec{U}_c d\ell
-{1\over 2}\int\int_A(\sigma_{ij}^g e_{ij}^c+\sigma_{ij}^c e_{ij}^q) dA .
\label{82}
\end{eqnarray}
The first two integrals in (\ref{82}) give the energy release for the
equilibrium elastic state with the specified cut opening $(g_x,g_y)$ and
tension $T$ at infinity. According to our decomposition this energy release
is maximum for given $g_x(x)$ and $g_y(x)$ and thus can not increase
linearly by tuning $\vec{U}_c$.  The latter is true only if the last two
integrals on the RHS of $(\ref{82})$ - linear in $\vec{U}_c$ - cancel each
other.  (This can be verified explicitly by integrating by parts the last
integral on the RHS of (\ref{82}) and using the fact that $\vec{U}_g$ is an
equilibrium state.) Thus
\begin{eqnarray}
E&=&\oint_{\Gamma_b} T \vec{n}(\vec{U}_g-\vec{U}_0) d\ell
+{1\over 2}\int\int_A (\sigma_{ij}^0 e_{ij}^0 -\sigma_{ij}^g e_{ij}^g) d A
\nonumber\\
&&-{1\over 2}\int\int_A \sigma_{ij}^c e_{ij}^c  d A :
\label{83}
\end{eqnarray}   
the energy factors, and the last term representing the 
continuous degrees of freedom  does not ``feel'' 
the presence of the cut and thus will have exactly the 
same spectrum as that of  the uncracked material.   

Although the elastic state $\vec{U}_g$ is an equilibrium one, the cut
boundary is in general stressed, and so we still have to modify the result
of the second section for the energy release (\ref{17}).  From the first
equation in (\ref{14}), the elastic energy of the uncracked material is
given by 
\begin{equation}
{1\over 2}\int\int_A \sigma_{ij}^0 e_{ij}^0  d A=
{1\over 2}\oint_{\Gamma_b}T\vec{n}\vec{U}_0 d\ell .
\label{sp1}
\end{equation} 
The elastic energy of the material with the cut (the second equation in
(\ref{14})) is modified to incorporate the stressed cut boundary
\begin{equation}
{1\over 2}\int\int_A \sigma_{ij}^g e_{ij}^g  d A=
{1\over 2}\oint_{\Gamma_b}T\vec{n}\vec{U}_g d\ell+{1\over 2}\oint_
{\Gamma_h}\vec{F}_h\vec{U}_g d\ell .
\label{sp2}
\end{equation}
The second integral in (\ref{sp2}) is over the cut boundary $\Gamma_h$ and
$\vec{F_h}$ is the force we have to apply to the cut boundary to insure its
displacements satisfy (\ref{78}-\ref{79}).  With this change, following
(\ref{83}-\ref{sp2}) we find that the energy release as given by (\ref{17})
decreases by $\delta E$
\begin{equation}
\delta E={1\over 2}\oint_{\Gamma_h}\vec{F_h}\vec{U}_g d\ell .
\label{84}
\end{equation}
In the spirit of the third section, the equilibrium elastic 
state $\vec{U}_g$ can be described by the analytical functions 
$\phi(z)$ and $\psi(z)$; $\phi_0(z)=\phi(z)-T z/2$ and $\psi(z)$ 
are   holomorphic in the extended 
complex $z$ plane excluding the straight cut and are constrained  
to provide  displacement discontinuity $(g_x,g_y)$. 
The energy release $E_g$ is then smaller than the one given by (\ref{60}) 
by $\delta E$
\begin{eqnarray}
E_g&=&-{{\pi T}\over{4\mu}}(1+\chi){\rm Re}[y_1^g]-\delta E\label{sh3}
\end{eqnarray}  
where $y_1^g$ is the $1/z$ coefficient in the expansion of $\psi(z)$ 
at infinity.

\begin{figure}
\centerline{
\psfig{figure=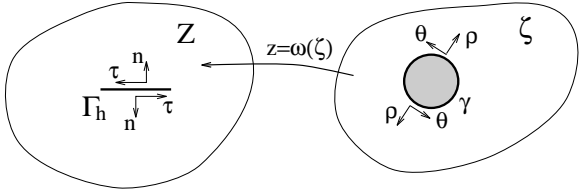,width=3truein}}
{\caption{The determination of the holomorphic functions 
describing the equilibrium elastic state of the material with 
a straight cut is simplified in the conformal plane $\zeta$,
where the unit circle $\gamma$ corresponds to cut boundary $\Gamma_h$ 
in the original $z$ plane.}
\label{f4}}
\end{figure}

To determine the functions $\phi_0(z)$ and $\psi(z)$ we conformally 
map the complex $z$ plane with the  cut  to the outside of the 
unit circle $\gamma$ (Figure \ref{f4}) 
\begin{equation}
z=\omega(\zeta)={{\ell}\over 4}\biggl(\zeta+{1\over{\zeta}}\biggr)
\label{85}
\end{equation}
so that the unit circle in the $\zeta$ plane is mapped to the straight 
cut boundary $\Gamma_h$ in the original plane, $\Gamma_h=\omega(\gamma)$.

The elasticity problem is reformulated in the conformal plane as follows:
we have to find analytical functions $\phi_g(\zeta)=\phi(\omega(\zeta))$ and
$\psi_g(\zeta)=\psi(\omega(\zeta))$, such that
${\phi_g}_0(\zeta)=\phi_g(\zeta)-\ell\zeta/4$ and $\psi_g(\zeta)$ are
holomorphic in the extended complex $\zeta$ plane outside the unit circle
and give the maximum energy release with displacement discontinuity
$(g_x,g_y)$ at the cut boundary.  We introduce 
\begin{equation}
g(\sigma)=u_g+iv_g\bigg|_{\gamma}
\label{86}
\end{equation}
where $\sigma=\exp(i\alpha)$, $\alpha\in[0,2\pi)$, is a parameterization  
of the  unit circle $\gamma$. 
Since $\sigma$ and $1/\sigma$ represent opposite points across the cut,
(\ref{78}-\ref{79}) require 
\begin{equation}
g(\sigma)-g(1/\sigma)=2[g_x(\omega(\sigma))+ig_y(\omega(\sigma))], 
\ \alpha\in[0,\pi) .
\label{sp4}
\end{equation} 
It is important to note that the equilibrium elastic state that maximizes 
the energy release for  given displacement discontinuity $(g_x,g_y)$ 
is unique; on the other hand (\ref{sp4}) determines only the 
asymmetric modes $g^{\rm asym}(\sigma)$  of the crack opening 
displacement for this state
\begin{eqnarray}
2 g^{\rm asym}(\sigma)&=&[u_g(\sigma)+iv_g(\sigma)]-
[u_g(1/\sigma)+iv_g(1/\sigma)]\label{sp5}\\
&=&2[g_x(\omega(\sigma))+ig_y(\omega(\sigma))] .\nonumber
\end{eqnarray}
The symmetric modes $g^{\rm sym}(\sigma)$ 
\begin{equation}
2 g^{\rm sym}(\sigma)=[u_g(\sigma)+iv_g(\sigma)]+
[u_g(1/\sigma)+iv_g(1/\sigma)]\label{sp6}
\end{equation}
left unconstrained by (\ref{sp4}), should then be relaxed 
to provide the maximum energy release for  given $g^{\rm asym}(\sigma)$.
Thus, to calculate the energy release $E_g$ of the elastic state
$\vec{U}_g$ we first find the energy release 
$E(g)=E(g^{\rm asym}+g^{\rm sym})$ for the equilibrium state with an 
arbitrary displacement along the cut boundary $g(\sigma)$ and then 
maximize the result with respect to $g^{\rm sym}$ 
\begin{equation}
E_g=\max\limits_{g^{\rm sym}} E(g^{\rm asym}+g^{\rm sym}) .
\label{sp7}
\end{equation}
In what follows we will use $\phi_{\zeta}(\zeta)$ and $\psi_{\zeta}(\zeta)$
to describe the equilibrium elastic state with an arbitrary 
displacement $g(\sigma)$ along the cut boundary and 
tension $T$ at infinity.
(The energy release for arbitrary $g(\sigma)$ is still given by 
(\ref{84}-\ref{sh3}).)
Making the change of variables $z\to w(\zeta)$
in (\ref{36}) and putting $\zeta=\sigma$ we obtain a constraint on 
$\phi_{\zeta}(\zeta)$ and $\psi_{\zeta}(\zeta)$ that guarantees 
the displacements along $\gamma$ to be $g(\sigma)$ 
\begin{equation}
\chi\phi_{\zeta}(\sigma)-{{\omega( \sigma)}\over{\overline{\omega'(\sigma)}}}
\overline{\phi_{\zeta}'(\sigma)}-\overline{\psi_{\zeta}(\sigma)}
=2\mu g(\sigma) .\label{88}
\end{equation}
Once the solution $(\phi_{\zeta},\psi_{\zeta})$ of the elasticity problem is
found, we can compute the correction (\ref{84}) to the energy release.
Introducing the polar coordinates $(\rho,\theta)$ (Figure \ref{f4}) in the
complex $\zeta$ plane, $\vec{F_h}=F_{\rho}\vec{\rho}+F_{\theta}\vec{\theta}$
and $\vec{U}_g=v_{\rho}
\vec{\rho}+v_{\theta}\vec{\theta}$, and using 
$d\ell=|\omega'(\sigma)||d\sigma|$ we find
\begin{eqnarray} 
\delta E&=&{1\over 2}\oint_{\gamma}\biggl(F_{\rho}v_{\rho}
+F_{\theta}v_{\theta}\biggr)|\omega'(\sigma)||d\sigma|
\label{89}\\
&=&{1\over 2}\oint_{\gamma}\biggl(\sigma_{\rho\rho}v_{\rho}
+\sigma_{\rho\theta}v_{\theta}\biggr)|\omega'(\sigma)||d\sigma|
\nonumber
\end{eqnarray}
where in the second equality we  express the force  
through the stress tensor components: $F_{\rho}=\sigma_{\rho\rho}$
and $F_{\theta}=\sigma_{\rho\theta}$.
The stress tensor components $\sigma_{\rho\rho}$ and 
$\sigma_{\rho\theta}$ are given in terms of the $\phi_{\zeta}(\zeta)$ and 
$\psi_{\zeta}(\zeta)$ functions that, as we already mentioned, completely 
determine the equilibrium elastic state. Muskhelishvili finds\cite{m}
\begin{eqnarray}
\sigma_{\rho\rho}-i\sigma_{\rho\theta}&=&
{{\phi_{\zeta}'(\zeta)}\over{\omega'(\zeta)}}
+{{\overline{\phi_{\zeta}'(\zeta)}}\over{\overline{\omega'(\zeta)}}}
-{{\zeta^2}\over{\rho^2
\overline{\omega'(\zeta)}}}\biggl\lbrace
{{\overline{\omega(\zeta)}}\over{\omega'(\zeta)}}
\phi_{\zeta}''(\zeta)\nonumber\\
&&-{{\overline{\omega(\zeta)} \omega''(\zeta)}
\over{{\omega'(\zeta)}}^2}\phi_{\zeta}'(\zeta)
+\psi_{\zeta}'(\zeta)\biggr\rbrace .
\label{90}
\end{eqnarray}
Noting that the transformation of  the displacements 
along the unit circle in the Cartesian coordinates $(u_g,v_g)$, 
$g=u_g+iv_g$, to the polar coordinates $(v_{\rho},v_{\theta})$ is\cite{m}
\begin{equation}
v_{\rho}+i v_{\theta}={1\over{\sigma}}{{\overline{\omega'(\sigma)}}
\over{|\omega'(\sigma)|}}(u_g+iv_g)={1\over{\sigma}}{{\overline{\omega'
(\sigma)}}\over{|\omega'(\sigma)|}}g(\sigma) ,
\label{91}
\end{equation}
we conclude from (\ref{89})
\begin{eqnarray}
\delta E&=&{1\over 2}{\rm Re}\oint_{\gamma}(\sigma_{\rho\rho}-
i\sigma_{\rho\theta})(v_{\rho}+i v_{\theta})|\omega'(\sigma)||d\sigma|
\label{92}\\
&=&{1\over 2}{\rm Re}\oint_{\gamma}(\sigma_{\rho\rho}-
i\sigma_{\rho\theta}){{\overline{\omega'(\sigma)}}\over{\sigma}} g(\sigma)
{{d\sigma}\over{i\sigma}}\nonumber
\end{eqnarray}
where $\sigma_{\rho\rho}-i\sigma_{\rho\theta}$ is given by (\ref{90})
with $\zeta\to\sigma$ ($\rho\to 1$).
From (\ref{sh3}) and (\ref{92}) we find the energy release $E(g)$ 
\begin{eqnarray}
E(g)&=&-{{\pi T}\over{4\mu}}(1+\chi){\rm Re} [y_1(g)]\nonumber\\
&&-{1\over 2}{\rm Re}\oint_{\gamma}(\sigma_{\rho\rho}-i\sigma_{\rho\theta}){
{\overline{\omega'(\sigma)}}\over{\sigma}} g(\sigma)
{{d\sigma}\over{i\sigma}}\label{93}
\end{eqnarray}
where $y_1(g)$ is the $1/z$ coefficient in the expansion 
of $\psi_{\zeta}(\omega^{-1}(z))$ at $z=\infty$.

The equilibrium elastic problem for material with the straight cut allows a
closed analytical solution for the arbitrary specified displacement
$g(\sigma)$ along the unit circle in the conformal plane $\zeta$.  Using the
fact that ${\phi_{\zeta}}_0(\zeta)$ and $\psi_{\zeta}(\zeta)$ are
holomorphic functions outside the unit circle that satisfy (\ref{88}),
Muskhelishvili finds\cite{m}
\begin{eqnarray}
\phi_{\zeta}(\zeta)&=&{{T\ell\zeta}\over 8}-{{2\mu}\over\chi}{1\over{2\pi i}}
\oint_{\gamma}{{g(\sigma) d\sigma}\over{\sigma-\zeta}}+{{T\ell}\over
{8\chi\zeta}}\label{94}\\
\psi_{\zeta}(\zeta)&=&{{\mu}\over{\pi i}}\oint_{\gamma}{{\overline{g(\sigma)}
d\sigma}\over{\sigma-\zeta}}+{{T\ell}\over 8}\biggl({\chi\over\zeta}-
{{2\zeta}\over{\zeta^2-1}}\biggr)\nonumber\\
&&-\zeta{{1+\zeta^2}\over{\zeta^2-1}}
\biggl(\phi_{\zeta}'(\zeta)-{{T\ell}\over 8}\biggr)-{{\mu}\over{\pi i}}
\oint_{\gamma}{{\overline{g(\sigma)}d\sigma}\over{\sigma}} .
\nonumber
\end{eqnarray}   
Assuming that $g(\sigma)$ is smooth, we represent it by a convergent 
Fourier series
\begin{equation}
g(\sigma)=\sum_{-\infty}^{+\infty}(a_n+ib_n)\sigma^n .\label{95}
\end{equation}
Using representation (\ref{95}) for $g(\sigma)$ we find from 
(\ref{93}) the energy release $E$
\begin{eqnarray}
E(g)&=&{{\pi T\ell(1+\chi)}\over{4\chi}}(
a_{-1}+\chi a_1)\label{97}\\
&&-2\pi\mu\sum_{n=1}^{+\infty}
n\biggl(a_n^2+b_n^2+{{a_{-n}^2+b_{-n}^2}\over\chi}\biggr)
\nonumber\\
&&-{{\pi T^2{\ell}^2(1+\chi)}\over{128\mu}}\biggl(\chi-2+
{1\over\chi}\biggr) .\nonumber
\end{eqnarray}
(The computations are tedious, but straightforward: first 
we substitute (\ref{95}) into (\ref{94}) to find the solution  
of the elasticity problem in terms of the Fourier amplitudes 
$\lbrace a_n,b_n\rbrace$, then we calculate the stress tensor 
components at the unit circle using (\ref{90}), and finally plugging 
the result into (\ref{93}) we obtain (\ref{97}).)
The next step is to relax the symmetric modes in the crack opening 
displacement given by $g(\sigma)$. 
From $(\ref{95})$  and $(\ref{86})$ we find
\begin{eqnarray}
u_g&=&\sum_{n=1}^{+\infty}(a_n+a_{-n})\cos n\alpha +(b_{-n}-b_n)\sin n\alpha
\label{z1}\\
v_g&=&\sum_{n=1}^{+\infty}(b_n+b_{-n})\cos n\alpha +(a_n-a_{-n})\sin n\alpha
\nonumber
\end{eqnarray} 
which with the change of variables 
\begin{eqnarray}
u_n&=&b_{-n}-b_n\label{z2}\\
v_n&=&a_n-a_{-n}\nonumber\\
\tilde{u}_n&=&b_n+b_{-n}\nonumber\\
\tilde{v}_n&=&a_n+a_{-n}\nonumber
\end{eqnarray} 
is rewritten as 
\begin{eqnarray}
u_g&=&\sum_{n=1}^{+\infty}\tilde{v}_n\cos n\alpha +u_n\sin n\alpha
\label{z3}\\
v_g&=&\sum_{n=1}^{+\infty}\tilde{u}_n\cos n\alpha +v_n\sin n\alpha .
\nonumber
\end{eqnarray}
It is clear now that the asymmetric modes of the crack opening 
displacement are described by $\lbrace u_n,v_n\rbrace$, while 
the symmetric ones are specified by $\lbrace \tilde{u}_n,\tilde{v}_n\rbrace$.
(Recall that points parameterized by $\sigma$ and $1/\sigma$
(or equivalently $\alpha$ and $-\alpha$) are opposite from one 
another across the cut.)
The amplitudes $\lbrace u_n,v_n\rbrace$ are uniquely determined 
for the given $(g_x,g_y)$. From (\ref{sp5}) and (\ref{z3})  
\begin{equation}
g_x(\ell/2\cos\alpha)+ig_y(\ell/2\cos\alpha)=
\sum_{n=1}^{+\infty}(u_n+iv_n)\sin n\alpha\label{z4}
\end{equation}
where $\alpha\in[0,\pi]$.
Using the  transformation inverse to (\ref{z2}) we can express the 
energy release (\ref{97}) in terms of 
$\lbrace u_n,v_n,\tilde{u}_n,\tilde{v}_n\rbrace$. The obtained  expression 
is maximum for 
\begin{eqnarray}
\tilde{u}_n&=&u_n{{\chi-1}\over{1+\chi}}\label{z5}\\
\tilde{v}_n&=&v_n{{1-\chi}\over{1+\chi}},
\mbox{\hspace{0.1in} }n\ne 1\nonumber\\
\tilde{v}_1&=&v_1{{1-\chi}\over{1+\chi}}+{{T\ell(\chi-1)}\over{8\mu}}
\nonumber
\end{eqnarray}
and gives the energy release $E_g$  
\begin{equation}
E_g={{T\ell\pi}\over 2}v_1 -{{2\pi\mu}\over{\chi+1}}\sum_{n=1}^{+\infty}
n\biggl(u_n^2+v_n^2\biggr) .\label{z6}
\end{equation}
Finally, the maximum of (\ref{z6}) is  achieved  for 
\begin{eqnarray}
u_n^{\rm max}&=&0\label{z7}\\
v_n^{\rm max}&=&0, \mbox{\hspace{0.1in} }n\ne 1\nonumber\\
v_1^{\rm max}&=&{{\pi T\ell(1+\chi)}\over{8\mu}}\nonumber
\end{eqnarray} 
and 
\begin{equation}
E_g^{\rm max}={{\pi T^2\ell^2(1+\chi)}\over{32\mu}}
\label{101}
\end{equation} 
which, as one might expect, corresponds to the equilibrium opening
of the  cut\cite{m} and the energy release associated 
with this opening (\ref{st}).
Expanding (\ref{z6}) about $\lbrace u_n^{\rm max},
v_n^{\rm max}\rbrace$, $\lbrace u_n,v_n\rbrace\to 
\lbrace u_n^{\rm max}+u_n,v_n^{\rm max}+v_n\rbrace$, 
we find
\begin{equation}
E_g={{\pi T^2\ell^2(1+\chi)}\over{32\mu}}-{{2\pi\mu}\over{1+\chi}}
\sum_{n=1}^{+\infty}n \biggl(u_n^2+v_n^2\biggr) .
\label{102}
\end{equation}
Expression (\ref{102}) is the desired result:
we find that the crack opening displacements (specified on the unit 
circle in the conformal plane)
\begin{eqnarray}
\lbrace u,v\rbrace=\bigg\lbrace &&v_n{{1-\chi}\over{1+\chi}}\cos n\alpha+
u_n\sin n\alpha,\nonumber\\
&& u_n{{\chi-1}\over{1+\chi}}\cos n\alpha+v_n\sin 
n\alpha \bigg\rbrace\label{103}
\end{eqnarray}
imposed on the saddle point cut opening 
\begin{equation}
\lbrace u^{\rm max},v^{\rm max}\rbrace=\lbrace 0,
{{\pi T\ell(1+\chi)}\over{8\mu}}\sin\alpha\rbrace\label{104}
\end{equation}
diagonalize the energy release 
and thus are the normal modes; with the excitation of the $n$-th normal
mode with the amplitude $\lbrace u_n,v_n\rbrace$ the energy release 
decreases by $2\pi\mu n (u_n^2+v_n^2)/(1+\chi)$.

Although (\ref{z6}) has been derived for the  material under 
 uniform isotropic stretching at infinity, it can be reinterpreted 
to describe the minimum increase  in the energy $\Delta E$ of the material  
under a uniform isotropic compression (pressure) $P$ at infinity, 
due to the opening of the straight cut with  specified displacement 
discontinuity along its boundary. For the displacement discontinuity 
given by (\ref{z4}) we find similar to (\ref{z6})
\begin{equation}
\Delta E={{P\ell\pi}\over 2}v_1 +{{2\pi\mu}\over{\chi+1}}\sum_{n=1}^{+\infty}
n\biggl(u_n^2+v_n^2\biggr) .\label{zz6}
\end{equation} 
One can use the same arguments that lead to (\ref{83}) to show that the crack 
opening normal modes (\ref{103}) decouple from all continuous 
modes (that are present in the uncracked material) and thus 
leave their spectrum unchanged. 
The saddle point is however unphysical in this case:
as follows from (\ref{104}), ($T$ in (\ref{104}) should be replaced with $-P$),
it corresponds to a configuration where the material overlaps itself.

\section{The imaginary part of the partition function}
Elastic materials at finite  temperature undergo a 
phase transition to fracture at zero applied stress, similar to 
the first order phase transition in spin systems below the critical 
temperature at zero magnetic field. 
The free energy of an elastic material under a stretching load develops 
an imaginary part which determines the material lifetime with respect to 
fracture. The imaginary part of the free energy has an essential singularity 
at zero applied stress. In this section we calculate this singularity 
at  low temperatures in a saddle point approximation including quadratic 
fluctuations. 

Consider an infinite two-dimensional  elastic material
subject to a uniform isotropic stretching $T$ at infinity.  Creation of a
straight cut of length $\ell$ will increase the energy 
by $2 \alpha \ell$, where $\alpha$ is the surface tension 
(the energy per unit length of edge), with a factor of $2$ because of 
the two free surfaces. On the other hand, the cut will open up because
of elastic relaxation.  Using (\ref{101}) for the energy release we find the 
total energy $E(\ell)$ of the straight cut in equilibrium under 
 stretching tension $T$:
\begin{equation} E(\ell)=2 \alpha \ell -{{\pi T^2\ell^2(1+\chi)} \over
{32\mu}} .
\label{105}
\end{equation}
Introducing 
\begin{equation}
\ell_c={{32\mu\alpha} \over {\pi T^2 (1+\chi)}}
\label{106}
\end{equation}
we can rewrite the energy of the crack as
\begin{equation}
E(\ell)=2\alpha \ell -\alpha {\ell^2 \over \ell_c} .
\label{107}
\end{equation}
It follows that cracks with $\ell>\ell_c$ will grow, giving rise to
the fracture of the material, while those with $\ell<\ell_c$ will heal
--- a result first obtained by Griffith\cite{g}. At finite temperature 
a crack of any size can appear as a thermal fluctuation, which 
means that for arbitrary small stretching $T$ the true ground 
state of the system is fractured into pieces and so the free 
energy of the material cannot be analytical at $T=0$.
Because the energy $E(\ell_c) = \alpha \ell_c$
grows as $1/ T^2$ as $T \rightarrow 0$, interactions between 
thermally nucleated cracks are unimportant at small $T$ and low temperatures
(allowing us to use the ``dilute gas approximation'').
 
The thermodynamic properties of a macroscopic system can be obtained 
from its partition function $Z$:
\begin{equation}
Z= \sum_{N=0}^{\infty}\sum_n \exp (-\beta E_{nN})\label{107a}
\end{equation}     
where the summation  $N$ is over all possible numbers of particles 
(cracks in our case) and the summation $n$ is over all states of the
system with $N$ cracks. 

To begin with, let's consider the partition function of  the material with
one cut $Z_1$ 
\begin{equation}
Z_1=\sum_E \exp (-\beta E)
\label{107c}
\end{equation}
where the summation is over all energy states of the material with a 
single cut. The calculation of the imaginary part of the partition function is
dominated by a saddle point, that in our case is a straight cut of 
length $\ell_c$.  The straight cut is the saddle point because 
it gains the most elastic relaxation energy for a given number of broken
bonds (we explicitly show in section III that curving  a cut 
reduces the energy release). 
For now we neglect all fluctuations of the critical droplet
(the cut of  length $\ell_c$) except for its uniform 
contraction or expansion --- fluctuations in the length of the straight 
cut.    Introducing the deviation $\Delta\ell$ in the cut 
length from the critical length $\ell_c$, $\Delta\ell=\ell-\ell_c$, 
we find from (\ref{107})
\begin{equation}
E=\alpha\ell_c -\alpha{{{\Delta\ell}^2}\over{\ell_c}} .
\label{107b}
\end{equation}
The fact that this degree of freedom has negative eigenvalue means that
direct computation of the partition function yields a divergent result.  A
similar problem for the three-dimensional Ising model was solved by
Langer\cite{langer}: one has to compute the partition function in a stable
state $P=-T$ (compression), and then do an analytical continuation in
parameter space to the state of interest. The free energy develops an
imaginary part in the unstable state, related to the decay rate for
fracture\cite{langer2}:  the situation is similar to that of barrier
tunneling in quantum mechanics \cite{affleck}, where the imaginary part in
the energy gives the decay rate of a resonance. We have explicitly
implemented this prescription for the simplified calculation of the
imaginary part of the free energy\cite{we}:  for the elastic material under
a uniform isotropic compression at infinity allowing for the nucleation of
straight cuts of an arbitrary length with an arbitrary elliptical opening
(mode $v_1$ in (\ref{zz6})), we calculated the free energy in a dilute gas
approximation.  We carefully performed the analytical continuation to the
metastable state describing the elastic material under the uniform isotropic
stretching $T$ at infinity and found the imaginary part of the free energy 
\begin{equation}
{\rm Im} F^{\rm simple}(T)={2 \over {\beta^2  T\lambda^2} }
\biggl ({ \pi {{A} \over {\lambda ^2}}} \biggr )\exp{ 
\biggl\lbrace  {{-32 \beta\mu
 \alpha^2 } \over { \pi T^2 (\chi+1)}  } \biggr\rbrace}
\label{w1}
\end{equation}  
where $A$ is the area of the material and $\lambda$ is the ultraviolet
cutoff of the theory. (The version of equation (\ref{w1})
as derived in\cite{we}, overcounts the contribution 
from zero-restoring-force modes $(2\pi A/\lambda^2)$ by factor $2$. 
Because cracks tilted  by $\theta$ and $\pi+\theta$  are identical, 
the proper contribution from rotations must be $\pi$, rather than $2\pi$.)

The alternative to this analytical continuation approach is to deform
 the integration contour over the amplitude of the unstable 
(negative eigenvalue) mode from the saddle point $\Delta\ell=0$ along 
the path of the steepest descent\cite{langer}. More precisely, we 
regularize the direct expression for the partition function 
\begin{equation}
Z_1=Z_0\biggl(\pi{A\over{\lambda^2}}\biggr)
\int_{-\ell_c}^{\infty}{{d\Delta\ell}\over\lambda}\exp\biggl\lbrace-\beta
\biggl(\alpha\ell_c -\alpha{{{\Delta\ell}^2}\over{\ell_c}}\biggr)\biggr
\rbrace\label{i1}
\end{equation} 
(which diverges at big $\Delta\ell$) 
by bending the $\Delta\ell$ integration contour 
from the saddle into the complex plane:
\begin{eqnarray}
Z_1&=&Z_0\biggl(\pi{A\over{\lambda^2}}\biggr)
\int_{-\ell_c}^{0}{{d\Delta\ell}\over\lambda}\exp\biggl\lbrace-\beta
\biggl(\alpha\ell_c -\alpha{{{\Delta\ell}^2}\over{\ell_c}}\biggr)\biggr
\rbrace\label{i2}\\
&&+Z_0\biggl(\pi{A\over{\lambda^2}}\biggr)\exp(-\beta\alpha\ell_c)
\int_{0}^{\pm i\infty}{{d\Delta\ell}\over\lambda}
\exp\biggl\lbrace\beta\alpha{{{\Delta\ell}^2}\over{\ell_c}}\biggl
\rbrace\nonumber
\end{eqnarray}  
In (\ref{i1}-\ref{i2}) the factor $( \pi A/{\lambda^2})$ comes  
from the zero-restoring-force modes for rotating and translating the cut,  
and $Z_0$ is the partition function for the uncracked material 
(unity for the present simplified calculation).
The second integral in (\ref{i2}) generates the imaginary part of the 
partition function
\begin{equation}
{\rm Im}Z_1=\pm{1\over 2}Z_0\biggl(\pi{A\over{\lambda^2}}
\biggr)\exp(-\beta\alpha\ell_c){\biggl({{\pi\ell_c}\over{
\beta\alpha\lambda^2}}\biggr)}^{1/2}
\label{i3}
\end{equation}
with the $\pm$ sign corresponding to the analytical continuation 
to either side of the branch cut of the partition function.
(We showed in\cite{we} that partition function is an analytical function 
in  complex $T$  with a branch cut along the line  $T\in[0,+\infty$).)
In a dilute gas approximation the partition function for the material
with $N$ cuts $Z_N$ is given by 
\begin{equation}
Z_N=Z_0{{{({Z_1}/ Z_0)}^N}\over{N!}}
\label{119}
\end{equation}  
which from (\ref{107a}) determines the material free 
energy 
\begin{equation}
F=-{1\over\beta}\ln Z=-{1\over\beta}\ln\sum_{N=0}^{\infty}Z_N =
-{1\over\beta}\ln Z_0-{1\over\beta}{{Z_1}\over{Z_0}} .
\label{120}
\end{equation}
Following (\ref{i3}) and (\ref{120}) we find the imaginary part 
of the free energy
\begin{eqnarray}
{\rm Im} F^{\rm simple}(T)=\pm&&{2 \over {\beta^2  T\lambda^2} }
{\biggl({{2\beta\mu\lambda^2}\over{\chi+1}}\biggr)}^{1/2}
\biggl ({ \pi {{A} \over {\lambda ^2}}} \biggr )\nonumber\\
&&\exp{ \biggl\lbrace  {{-32 \beta\mu
 \alpha^2 } \over { \pi T^2 (\chi+1)}  } \biggr\rbrace}
\label{i4}
\end{eqnarray} 
(\ref{i4}) differs from (\ref{w1}) only because for the calculation of the
imaginary part of the free energy in\cite{we} we used two degrees of
freedom: the length of the cut and its elliptical opening, while in current
calculation there is only one degree of freedom. One can immediately restore
(\ref{w1}) by adding the $v_1$ mode of (\ref{102}) to the energy of the
elastic material (\ref{107b}) and integrating it out. From (\ref{102}), the
$v_1$ mode generates an additional multiplicative contribution $Z_{v_1}$ to
the partition function for a single crack $Z_1$, and thus from (\ref{120})
changes the imaginary part of the free energy for multiple cracks $F^{\rm
simple}$, ${\rm Im}F^{\rm simple}\to Z_{v_1} {\rm Im}F^{\rm simple}$ 
\begin{equation}
Z_{v_1}=\int_{-\infty}^{+\infty}{{dv_1}\over{\lambda}}\exp\biggl\lbrace
-{{2\pi\mu\beta}\over{1+\chi}}v_1^2\biggr\rbrace={\biggl({{1+\chi}
\over{2\mu\beta\lambda^2}}\biggr)}^{1/2}
\label{ii4}
\end{equation}
which will cure the discrepancy between (\ref{i4}) and (\ref{w1}).
Although the analytical continuation method is theoretically 
more appealing, the calculation of the imaginary part through 
the deformation of the integration contour of the unstable mode 
is more convenient once we include the quadratic fluctuations.
It is clear that both methods (properly implemented) must give the 
same results.

We have already emphasized that the above calculation ignores 
the quadratic fluctuations about the saddle point (except for the uniform 
contraction or extension of the critical droplet), 
which may change the prefactor in the expression (\ref{i4}) for the 
imaginary part of the free energy and may renormalize the surface tension
$\alpha$. There are three kinds of quadratic fluctuations we have to deal
with. (I) {\it Curvy cuts} --- changes in the shape of the tear in the 
material: deviations of the broken bonds from a straight-line configuration.
(II) {\it Surface phonons} --- thermal fluctuations of the free surface 
of the crack about its equilibrium opening. (III) {\it Bulk phonons} ---
thermal fluctuations of the elastic media that are continuous at the cut 
boundary. To incorporate these fluctuations we have to integrate out the 
quadratic deviation from the saddle point energy coming from their degrees
of freedom (as we did for the surface phonon $v_1$ above).
In all cases the answer will depend upon the microscopic lattice-scale 
structure of the material. In field-theory language, our
theory needs regularization: we must decide exactly how
to introduce the ultraviolet cut-off $\lambda$. Here we discuss the lattice 
regularization, where the cut-off is explicitly introduced by the 
interatomic distance, and  $\zeta$-function 
regularization, common in field theory. We find that  the precise form 
of the surface tension 
renormalization and the prefactor in the imaginary part of the free energy 
depends on the regularization prescription, but certain important quantities 
appear regularization independent.

The partition function of the elastic material with one cut 
$Z_1$ in the saddle point approximation (\ref{i2}),
will develop a multiplicative factor $Z_f$ upon inclusion of the quadratic 
fluctuations $Z_1\to Z_f Z_1$ with 
\begin{equation}
Z_f=\sum_{\Delta E}\exp(-\beta \Delta E) .
\label{131}
\end{equation}
A deviation $\Delta E$ from the saddle point energy is decomposed 
into three parts, with each part describing  fluctuations of one 
of the mentioned three types
\begin{eqnarray}
\Delta E&=&{{\alpha\l_c^2}\over{\pi^2\lambda}}\sum_{n=1}^{\infty}{1\over{n^2}}
c_n^2+{{2\pi\mu}\over{1+\chi}}\sum_{n=1}^{\infty}n(u_n^2+v_n^2)\nonumber\\
&&+\Delta E_{\rm continuous} .\label{132}
\end{eqnarray} 
The first term in (\ref{132}) accounts for the decrease in the 
energy release due to the curving of the saddle point cut of 
length $\ell_c$ with the curvature
\begin{equation}
k(x)=\sum_{n=1}^{\infty}c_n\sqrt{{{2\lambda}\over{\ell_c}}}\sin
{{\pi n x}\over{\ell_c}},\ x\in[0,\ell_c] .
\label{133}
\end{equation}
(The first term in (\ref{132}) follows from (\ref{c9}) with $l=l_c$ 
given by (\ref{106}).)   
The second term in (\ref{132}) describes the asymmetric modes in the 
thermal fluctuations of the free surface of the saddle point crack
about its equilibrium opening shape 
\begin{equation}
u^{\rm asym}(t)+i v^{\rm asym}(t)=
\sum_{n=1}^{\infty}(u_n+iv_n)\sin n\vartheta,\ \vartheta\in[-\pi,\pi)
\label{134}
\end{equation} 
where a point at the cut boundary is parameterized by its 
distance $t=\ell_c(1+\cos\vartheta)/2$ from the cut end;
$\vartheta\in[-\pi,0)$ parameterize the lower boundary displacements
and $\vartheta\in[0,\pi)$  parameterize the displacements of the 
upper boundary points. The symmetric modes of the crack 
opening about its equilibrium opening shape are assumed to relax 
providing the minimum increase in the elastic energy for a given 
$\lbrace u_n,v_n\rbrace$. The latter guarantees that all additional 
modes with the continuous displacement at the cut boundary 
(the ones which give $\Delta E_{\rm continuous}$ --- the last 
term in (\ref{132}) describing the bulk phonons) 
decouple from $\lbrace c_n,u_n,v_n\rbrace$ 
and are the same as the ones for the uncracked material. 
(The arguments here are the same as those that were used in derivation of 
(\ref{83}).) Since the curvature modes $\lbrace c_n\rbrace$ give the 
equilibrium energy of the curvy cut, the response of the surface 
phonons to such a curving is already incorporated, so 
the quadratic fluctuations $\lbrace c_n\rbrace$ can be calculated 
independently from the quadratic fluctuations $\lbrace u_n,v_n\rbrace$.
The latter means that there are no coupling between $\lbrace c_n\rbrace$
and $\lbrace u_n,v_n\rbrace$ modes in (\ref{132}), and the 
spectrum of $\lbrace u_n,v_n\rbrace$ modes is the same as that for
the straight cut of length $\ell_c$ (\ref{102}). 

The last thing we have to settle before the calculation of $Z_f$
is the proper integration measure for the surface phonon modes 
$\lbrace u_n,v_n\rbrace$. (We argued in the conclusion of  section 
 III that the integration measure for the modes $c_n$ is Cartesian --- 
$\prod_{n=1}^{\infty}dc_n$.) Here we show that because the functional measure 
in the displacement fields $(u(x,y),v(x,y))$ defined at each point 
of the material $(x,y)$ is naturally Cartesian --- 
$D[u(x,y)/\lambda]D[v(x,y)/\lambda]$,
the integration measure for the modes $\lbrace u_n,v_n\rbrace$
must be of the form 
$\prod_{n=1}^{\infty}(1/2\pi)du_n dv_n/\lambda^2$.

An arbitrary elastic displacement field for the material with a curvy cut 
is defined by specifying its bulk part $(u_{\rm bulk}(x,y),v_{\rm bulk}(x,y))$ 
(point (x,y) can be anywhere except at the cut boundary) and the cut part 
$(u_{\rm cut}^+(t),u_{\rm cut}^-(t),v_{\rm cut}^+(t),v_{\rm cut}^-(t))$ 
(the cut displacements are defined along the cut and are parameterized 
by the distance $t=\ell_c(1+\cos\vartheta)/2,\ \vartheta\in[0,\pi)$ 
from the cut end; the $+$ and $-$ superscripts are correspondingly 
the displacements at the upper and the lower boundary of the cut).
It is helpful to visualize the introduction of the cut into the material 
as splitting in half each of the  atoms of the material along the 
cut boundary. 
Then, the bulk part of the displacement field combines degrees of 
freedom of all atoms left untouched by splitting and the cut part 
describes the displacements of the split ones. Note that the splitting
increases the total number of the degrees of freedom.
The original measure is naturally 
\begin{eqnarray}
D[u_{\rm bulk}(x,y)/\lambda]&& 
D[v_{\rm bulk}(x,y)/\lambda]D[u_{\rm cut}(t)^+u_{\rm cut}(t)^-/\lambda^2]
\nonumber\\
&&D[v_{\rm cut}(t)^+v_{\rm cut}(t)^-/\lambda^2]\nonumber
\end{eqnarray}
First we separate the symmetric and asymmetric parts in the crack 
opening displacement
\begin{eqnarray}
u^{\rm asym}(t)&=&{1\over 2}\biggl(u_{\rm cut}^+(t)-u_{\rm cut}^-(t)\biggr)
\label{135}\\
v^{\rm asym}(t)&=&{1\over 2}\biggl(v_{\rm cut}^+(t)-v_{\rm cut}^-(t)\biggr)
\nonumber\\
u^{\rm sym}(t)&=&{1\over 2}\biggl(u_{\rm cut}^+(t)+u_{\rm cut}^-(t)
\biggr)\nonumber\\
v^{\rm sym}(t)&=&{1\over 2}\biggl(v_{\rm cut}^+(t)+v_{\rm cut}^-(t)
\biggr)\nonumber
\end{eqnarray}
Because the Jacobian of the transformation 
\begin{eqnarray}
&&(u_{\rm cut}(t)^+,u_{\rm cut}(t)^-,v_{\rm cut}(t)^+,v_{\rm cut}(t)^-)
\to\nonumber\\
&&(u^{\rm asym}(t),v^{\rm asym}(t),u^{\rm sym}(t),v^{\rm sym}(t))\nonumber 
\end{eqnarray}
is constant
\begin{equation}
\Bigg|{{\partial(u_{\rm cut}(t)^+,
u_{\rm cut}(t)^-,v_{\rm cut}(t)^+,
v_{\rm cut}(t)^-)}\over{\partial(u^{\rm asym}(t),v^{\rm asym}(t),
u^{\rm sym}(t),v^{\rm sym}(t))}}\Bigg|={1\over 4} ,
\label{136}
\end{equation}
the integration measure remains Cartesian: 
\begin{eqnarray}
D[u_{\rm bulk}(x,y)/\lambda] 
&&D[v_{\rm bulk}(x,y)/\lambda]D[u^{\rm sym}(t)v^{\rm sym}(t)/\lambda^2]
\nonumber\\&&D[u^{\rm asym}(t)v^{\rm asym}(t)/4\lambda^2] .\nonumber
\end{eqnarray}
Now we can combine the bulk and the symmetric cut part of the measure
by introducing the continuous displacement fields $(u_c(x,y),v_c(x,y))$ 
everywhere, including the cut boundary.(In our atomic picture, the symmetric 
modes of the cut part of the displacement fields represent the displacements
of the split atoms as if they were whole, and so it is natural 
to combine these degrees of freedom with the bulk ones.  Obtained as 
a result of such combination the continuous degrees of freedom are 
indistinguishable from the degrees of freedom of the uncracked material.) 
The integration measure 
becomes 
$$D[u_c(x,y)/\lambda]D[v_c(x,y)/\lambda]D[u^{\rm asym}(t)
v^{\rm asym}(t)/4\lambda^2] .$$ According to our decomposition, 
we specify the asymmetric cut opening and 
find the equilibrium displacement fields that minimize the increase 
in the elastic energy. In other words,  given
$(u^{\rm asym}(t),v^{\rm asym}(t))$ determine 
$(u_c^{\rm min}(u^{\rm asym},v^{\rm asym}),v_c^{\rm min}(u^{\rm asym},
v^{\rm asym}))$. The transformation
\begin{eqnarray}
u_c(x,y)&=&u_c^{\rm min}(u^{\rm asym},v^{\rm asym})+\tilde{u}_c(x,y)
\label{137}\\
v_c(x,y)&=&v_c^{\rm min}(u^{\rm asym},v^{\rm asym})+\tilde{v}_c(x,y)
\nonumber
\end{eqnarray} 
then completely decouple the surface phonon modes and the continuous modes 
that contribute to $\Delta E_{\rm continuous}$ in (\ref{132}). 
The Jacobian of the transformation 
\begin{eqnarray}
&&(u_c(x,y),v_c(x,y),u^{\rm asym}(t),
v^{\rm asym}(t))\to\nonumber\\&&(\tilde{u}_c(x,y),\tilde{v}_c(x,y),
u^{\rm asym}(t),v^{\rm asym}(t))\nonumber
\end{eqnarray}
 is unity (the transformation is 
just a functional shift) and so the measure remains unchanged
$$D[\tilde{u}_c(x,y)/\lambda]D[\tilde{v}_c(x,y)/\lambda]D[u^{\rm asym}(t)
v^{\rm asym}(t)/4\lambda^2] .$$
The Fourier transformation (\ref{134}) is orthogonal, but the Fourier 
modes are not normalized:
\begin{equation}
\int_0^{\pi}d\vartheta\ \sin^2 n\vartheta ={\pi\over 2} .
\label{137a}
\end{equation}
The latter means that at the final stage of the change of variables
 $(u^{\rm asym}(t),v^{\rm asym}(t))\to \lbrace u_n,v_n\rbrace$
there appear the Jacobian $\prod_{n=1}^{\infty}(2/\pi)$, and so we end up
with the integration measure 
$\prod_{n=1}^{\infty}(1/2\pi)du_n dv_n/\lambda^2$.

From (\ref{131}-\ref{132}) with the proper integration measure over 
the surface phonon modes we find
\begin{eqnarray}
Z_f&=&\prod_{n=1}^{\infty}\int_{-\infty}^{+\infty}dc_n
\exp\biggl\lbrace -\beta{{\alpha\l_c^2}\over{\pi^2\lambda n^2}}c_n^2
\biggr\rbrace\label{145}\\
&&\prod_{n=1}^{\infty}{\int\int}_{-\infty}^{+\infty}
{1\over{ 2\pi}}{{du_n dv_n}\over{\lambda^2}}
\exp\biggl\lbrace -\beta{{2\pi\mu n}\over{1+\chi}}(u_n^2+v_n^2)
\biggr\rbrace\nonumber\\
&&Z_{\rm continuous}\nonumber\\
&=&\prod_{n=1}^{\infty}{\biggl({{\pi^3\lambda n^2}\over{\beta\alpha
\ell_c^2}}\biggr)}^{1/2}\ \prod_{n=1}^{\infty}{{1+\chi}\over
{4\pi\beta\mu\lambda^2 n}}\  Z_{\rm continuous}
\label{145a}
\end{eqnarray}
where 
\begin{equation}
Z_{\rm continuous}=\sum_{\Delta E_{\rm continuous}}
\exp(-\beta\Delta E_{\rm continuous}) .
\label{146}
\end{equation}
Because $\Delta E_{\rm continuous}$ corresponds to the degrees of freedom 
of the uncracked material (with the same energy spectrum),
 $Z_{\rm continuous}$ contributes to the partition function $Z_0$ of 
the material without the crack, which according to (\ref{120}) drops out 
from the calculation of the imaginary part of the free energy. 

All the products over $n$ in these expressions diverge: we need a
prescription for cutting off the modes at short wavelengths (an ultraviolet
cutoff).

First we'll consider the $\zeta$-function regularization.
In this  regularization prescription\cite{ramond},
the infinite product of the type  $D=\prod_{n=1}^{\infty}\lambda_n$ 
is evaluated by introducing  the function $D_{\zeta}(s)$ 
\begin{equation}
D_{\zeta}(s)=\sum_{n=1}^{\infty}{1\over{\lambda_n^s}}
\label{151}
\end{equation}
so that 
\begin{equation}
D=\exp(-D_{\zeta}'(0)) .
\label{152}
\end{equation}
It is assumed that the sum (\ref{151}) is convergent in some region of the 
complex  $s$ plane and that it is possible to analytically continue 
$D_{\zeta}(s)$ from that region to $s=0$. 
From (\ref{145a}) we find
\begin{equation}
Z_f=D_1 D_2 Z_{\rm continuous}
\label{153}
\end{equation}
where $D_1$ and $D_2$ are obtained following (\ref{152}) 
from the corresponding $\zeta$-functions: ${D_1}_{\zeta}(s)$ 
and ${D_2}_{\zeta}(s)$
\begin{eqnarray}
{D_1}_{\zeta}(s)&=&{\biggl({{\beta\alpha\ell_c^2}\over{\pi^3\lambda}}
\biggr)}^{s/2}\sum_{n=1}^{\infty}{1\over{n^s}}=
{\biggl({{\beta\alpha\ell_c^2}\over{\pi^3\lambda}}
\biggr)}^{s/2}\zeta_R(s)\label{154}\\
{D_2}_{\zeta}(s)&=&{\biggl({{4\pi\beta\mu\lambda^2}\over{1+\chi}}
\biggr)}^s\sum_{n=1}^{\infty}n^s={\biggl({{4\pi\beta\mu\lambda^2}\over{1+\chi}}
\biggr)}^s\zeta_R(-s) .\nonumber
\end{eqnarray}
$\zeta_R(s)$ in (\ref{154}) is the standard Riemann $\zeta$-function, 
 holomorphic everywhere in the complex $s$ plane except 
at  $s=1$. Noting that $\zeta(0)=-1/2$ and
$\zeta'(0)=-(\ln 2\pi)/2$ we find from (\ref{152}-\ref{154}) 
\begin{equation}
Z_f={\biggl({{4\beta\alpha\ell_c^2}\over
{\pi\lambda}}\biggr)}^{1/4}{\biggl({{2\beta\mu\lambda^2}\over
{1+\chi}}\biggr)}^{1/2} Z_{\rm continuous} .\label{155}
\end{equation} 
From (\ref{120}) and (\ref{145a}) we find the imaginary part of the 
free energy in the $\zeta$-function regularization 
\begin{equation}
{\rm Im}F^{\zeta}={\biggl({{16\beta^3\alpha\mu^2\lambda^5}\over
{\pi{(1+\chi)}^2}}\biggr)}^{1/4}{\biggl({{\ell_c}\over{\lambda}}\biggr)}
^{1/2}{\rm Im}F^{\rm simple}
\label{156}
\end{equation}
where ${\rm Im}F^{\rm simple}$ is given by (\ref{i4}).

Second, we consider  lattice regularization --- which is more elaborate. 
We represent a curvy cut by $N+1=\ell_c/\lambda$ segments of equal 
length parameterized by the
kink angles $\lbrace\alpha_i\rbrace,\ i\in[1,N]$. With our conventional 
parameterization of the cut $t=\ell_c(1+\cos\vartheta)/2,\ 
\vartheta[-\pi,\pi)$,
the asymmetric modes of the crack opening displacements 
$\lbrace u^{\rm asym}(t), v^{\rm asym}(t)\rbrace$ are linear 
piecewise approximation for  given 
asymmetric displacements of the ``split'' kink atoms $\lbrace u^{\rm asym}_i,
v^{\rm asym}_i\rbrace,\ i\in[1,N]$. More precisely, if $t_i$ and $t_{i+1}$
parameterize the adjacent kinks, we assume 
\begin{eqnarray}
u^{\rm asym}(t)&=&u^{\rm asym}_i+{{u^{\rm asym}_{i+1}-u^{\rm asym}_i}
\over{t_{i+1}-t_i}}(t-t_i)
\label{152a}\\
v^{\rm asym}(t)&=&v^{\rm asym}_i+{{v^{\rm asym}_{i+1}-v^{\rm asym}_i}
\over{t_{i+1}-t_i}}(t-t_i)\nonumber
\end{eqnarray} 
for $t\in[t_i,t_{i+1}]$.
From the integration measure arguments for the  
$\zeta$-function regularization, it is clear that the  integration 
measure in this case must be $\prod_{i=1}^N d\alpha_i \prod_{i=1}^{N}du^
{\rm asym}_i dv^{\rm asym}_i/4\lambda^2$.
Now we have to write down the lattice regularization of the 
quadratic deviation $\Delta E$ from the saddle point energy (\ref{132}).
From (\ref{c1}) and (\ref{c3}), the curving of the critical cut $\ell_c$ will 
reduce the energy release by $\Delta E_c$
\begin{eqnarray}
\Delta E_c &=&{{T^2\pi(1+\chi)\ell_c^2}\over{32\mu}}\sum_{i,j=1}^N
\alpha_i\alpha_j M^c_{ij}\nonumber\\ 
&=&\alpha\ell_c\sum_{i,j=1}^N\alpha_i\alpha_j M^c_{ij}
\label{153a}
\end{eqnarray} 
where 
\begin{equation}
M^c_{ij}=-{{i j}\over{{(N+1)}^2}}+{{{\rm min}(i,j)}\over{N+1}} .
\label{154a}
\end{equation}
From (\ref{132}) the surface phonon contribution to $\Delta E$ is given by
\begin{equation}
\Delta E_p={{2\pi\mu}\over{1+\chi}}\sum_{n=1}^{+\infty}n(u_n^2+v_n^2)
\label{155a}
\end{equation}  
where from (\ref{134})
\begin{eqnarray}
u_n&=&{2\over\pi}\int_0^{\pi}d\vartheta\ u^{\rm asym}(t(\vartheta))
\sin n\vartheta\label{156a}\\
v_n&=&{2\over\pi}\int_0^{\pi}d\vartheta\ v^{\rm asym}(t(\vartheta)) 
\sin n\vartheta .\nonumber
\end{eqnarray}
In principle, for a given piecewise approximation of the asymmetric
modes (\ref{152a}) determined by $\lbrace u^{\rm asym}_i,
v_i^{\rm asym}\rbrace$, $i\in[1,N]$, one could calculate the Fourier 
amplitudes according to (\ref{156a}) and then plug the result into 
(\ref{155a}) to obtain $\Delta E_p$ in terms of 
$\lbrace u^{\rm asym}_i,v_i^{\rm asym}\rbrace$. We will use another 
approach. Using $ u^{\rm asym}(0)=u^{\rm asym}(\ell_c)=
0$ (with the same equalities for $v^{\rm asym}(t)$) we integrate
(\ref{156a}) by parts to obtain
\begin{eqnarray}
u_n&=&{2\over\pi}\int_0^{\pi}d\vartheta\ {{d u^{\rm asym}(t(\vartheta))
}\over{d\vartheta}}
{{\cos n\vartheta}\over n}\label{157a}\\
v_n&=&{2\over\pi}\int_0^{\pi}d\vartheta\ {{d v^{\rm asym}(t(\vartheta))
}\over{d\vartheta}}{{\cos n\vartheta}\over n} .\nonumber
\end{eqnarray}   
Substituting (\ref{157a}) into (\ref{155a}) we find
\begin{eqnarray}
\Delta E_p&=&{{8\mu}\over{\pi(1+\chi)}}\int_0^{\pi}\int_0^{\pi}
d\vartheta_1 d\vartheta_2 \biggl[{{d u^{\rm asym}(\vartheta_1)}
\over{d\vartheta_1}}{{d u^{\rm asym}(\vartheta_2)}\over{d\vartheta_2}}
\nonumber\\
&&+{{d v^{\rm asym}(\vartheta_1)}\over{d\vartheta_1}}
{{d v^{\rm asym}(\vartheta_2)
}\over{d\vartheta_2}}\biggr]K(\vartheta_1,\vartheta_2)
\label{158}
\end{eqnarray}
where
\begin{eqnarray}
K(\vartheta_1,\vartheta_2)&=&\sum_{n=1}^{\infty}{{\cos n\vartheta_1 \cos 
n\vartheta_2}\over n} .
\label{158a}
\end{eqnarray}
Following\cite{rushic}
\begin{equation}
\sum_{k=1}^{\infty}{{\cos k x}\over{k}}={1\over 2}\ln{1\over{2(1-\cos x)}} ,
\label{159a}
\end{equation}
we find an analytical expression for the kernel (\ref{158a}) 
\begin{equation}
K(\vartheta_1,\vartheta_2)=-{1\over 2}\ln 2-{1\over 2}\ln|\cos\vartheta_1-
\cos\vartheta_2| .
\label{159}
\end{equation}
Finally, introducing  $M_{ij}^p$ from 
\begin{eqnarray}
\sum_{i,j=1}^N&& \biggl(u^{\rm asym}_i u^{\rm asym}_j+
v^{\rm asym}_i v^{\rm asym}_j \biggr) M_{ij}^p\nonumber\\
&=&\int_0^{\pi}\int_0^{\pi}
d\vartheta_1 d\vartheta_2 \biggl[{{d u^{\rm asym}(\vartheta_1)}\over
{d\vartheta_1}}{{d u^{\rm asym}(\vartheta_2)}\over{d\vartheta_2}}\nonumber\\
&&+{{d v^{\rm asym}(\vartheta_1)}\over{d\vartheta_1}}
{{d v^{\rm asym}(\vartheta_2)}\over{d\vartheta_2}}
\biggr]K(\vartheta_1,\vartheta_2)\label{160}
\end{eqnarray}
we obtain  
\begin{equation}
\Delta E_p={{8\mu}\over{\pi(1+\chi)}}\sum_{i,j=1}^N \biggl(
u^{\rm asym}_i u^{\rm asym}_j+ v^{\rm asym}_i v^{\rm asym}_j 
\biggr) M_{ij}^p .
\label{161}
\end{equation}
To calculate  $M_{ij}^p$ we  substitute (\ref{152a}) directly into the 
RHS of (\ref{160}) and read off the corresponding coefficient,
given by the following three equations: 
\begin{eqnarray}
&&M_{ij}^p=f_2(i,j)+f_2(i+1,j+1)-f_2(i+1,j)\nonumber\\
&&\qquad \quad -f_2(i,j+1)\nonumber\\\nonumber\\ 
&&f_2(i,j)={3\over 4}+{1\over{(\cos\vartheta_{i-1}-\cos\vartheta_i)
(\cos\vartheta_{j-1}-\cos\vartheta_j)}}\biggl[\nonumber\\
&&f_1(i,j)+f_1(i-1,j-1)-f_1(i-1,j)-f_1(i,j-1)\biggr]\nonumber\\
\nonumber\\&&f_1(i,j)=\nonumber\\
&&\cases{{{{(\cos\vartheta_i-\cos\vartheta_j)}^2}\over 4}
\ln\bigg|\sin{{\vartheta_i-\vartheta_j}\over 2}
\sin{{\vartheta_i+\vartheta_j}\over 2}\bigg|,&if $i\ne j$;\cr
0,&otherwise,\cr}\nonumber\\
\label{162}
\end{eqnarray}
where $\vartheta_0=0$, $\vartheta_{N+1}=\pi$, and $\vartheta_i$ 
parameterizes the $i$-th kink (kinks are equally spaced in real space):
\begin{equation}
\vartheta_i=\arccos\biggl(1-{{2 i}\over{N+1}}\biggr),\ i\in[1,N] .
\end{equation}
From (\ref{153a}) and (\ref{161}) we find the quadratic deviation 
from the saddle point energy 
\begin{eqnarray}
\Delta E &=& \Delta E_c +\Delta E_p +\Delta E_{\rm continuous}
\label{164}\\
&=&\alpha\ell_c\sum_{i,j=1}^N\alpha_i\alpha_j M^c_{ij}+
{{8\mu}\over{\pi(1+\chi)}}\sum_{i,j=1}^N \biggl(
u^{\rm asym}_i u^{\rm asym}_j\nonumber\\
&&+ v^{\rm asym}_i v^{\rm asym}_j 
\biggr) M_{ij}^p+\Delta E_{\rm continuous}.\nonumber
\end{eqnarray}
Thus the multiplicative factor $Z_f$ to the partition function of 
the elastic material with  one cut in the lattice regularization 
is given by
\begin{eqnarray}
Z_f&=&\prod_{n=1}^N\int_{-\infty}^{+\infty}d\alpha_n\exp
\biggl\lbrace -\beta\alpha\ell_c\sum_{i,j=1}^N 
\alpha_i\alpha_j M^c_{ij}\biggr\rbrace\label{165}\\
&&\prod_{n=1}^N\int\int_{-\infty}^{+\infty}{{du_n^{\rm asym}
dv_n^{\rm asym}}\over{4\lambda^2}}\exp\biggl\lbrace -\beta
{{8\mu}\over{\pi(1+\chi)}}\nonumber\\
&&\sum_{i,j=1}^N \biggl(
u^{\rm asym}_i u^{\rm asym}_j+ v^{\rm asym}_i v^{\rm asym}_j 
\biggr) M_{ij}^p\biggr\rbrace\nonumber\\
&& Z_{\rm continuous}\nonumber\\
&=& {\biggl({{\pi}\over{\beta\alpha\ell_c}}\biggr)}^{N/2}
{\det}^{-1/2} M_{ij}^c\ {\biggl({{\pi^2 (1+\chi)}\over{32\beta\mu\lambda^2}}
\biggr)}^N\nonumber\\ &&{\det}^{-1} M_{ij}^p\ Z_{\rm continuous}\nonumber
\end{eqnarray}
where $Z_{\rm continuous}$ is given by (\ref{146}).

The determinant coming from the curvy cuts $M_{ij}^c$ can be calculated
analytically. In section III we show that $\sin \pi n x/\ell$ are 
eigenvectors of the operator (\ref{c5}), the continuous analog of $M_{ij}^c$.
One can explicitly check that for $n\in[1,N]$, vectors  $\vec{m}_n=
\lbrace\sin\pi n i/(N+1)\rbrace$ are in fact eigenvectors of $M_{ij}^c$
with  eigenvalues
\begin{equation}
\lambda_n={1\over{4(N+1)}}\ \sin^{-2} {{\pi n}\over{2(N+1)}}
\label{166}
\end{equation}
and so
\begin{equation}
\det M_{ij}^c =\prod_{n=1}^N\lambda_n={\biggl(N+1\biggr)}^{-(N+1)} .
\label{167}
\end{equation}
(To obtain (\ref{167}), we take the limit $x\to 0$ of
\begin{equation}
\sin 2(N+1) x=2^{2 N+1} \prod_{k=0}^{2 N+1}\sin\biggl(x+{{k\pi}\over{2(N+1)
}}\biggr)\label{167a}
\end{equation}
\cite{rushic}, to get 
\begin{equation}
N+1=4^N\prod_{k=1}^{2 N+1}\sin{{k\pi}\over{2(N+1)
}}=4^N\prod_{k=1}^N\sin^2{{k\pi}\over{2(N+1)}} .\label{167b}
\end{equation}
With (\ref{167b}), the calculation in (\ref{167}) becomes straightforward.)
Recalling that $N+1=\ell_c/\lambda$, we can rewrite (\ref{165}) 
making use of (\ref{167})
\begin{eqnarray}
Z_f&=&\sqrt{{\beta\alpha\lambda}\over\pi}{\biggl({\pi\over{\beta\alpha\lambda}}
\biggr)}^{\ell_c/2\lambda}{\biggl({{\ell_c}\over\lambda}\biggr)}^{1/2}\
\sqrt{{32\beta\mu\lambda^2}\over{\pi^2 (1+\chi)}}\nonumber\\
&&{\biggl({{\pi^2 (1+\chi)}\over{32\beta\mu\lambda^2}}
\biggr)}^{\ell_c/2\lambda}{\det}^{-1} M_{ij}^p\  Z_{\rm continuous} .
\label{168}
\end{eqnarray}
Note that the first three factors on the RHS of (\ref{168})
(coming from the curvy cut fluctuations)  have the asymptotic form, 
$N\to\infty$, 
\begin{equation}
\sqrt{{\beta\alpha\lambda}\over\pi}{\biggl({\pi\over{\beta\alpha\lambda}}
\biggr)}^{\ell_c/2\lambda}{\biggl({{\ell_c}\over\lambda}\biggr)}^{1/2}
\approx N^{c_2} \exp\lbrace c_0+c_1 N\rbrace
\label{168a}
\end{equation}   
with $c_0=0$, $c_1=\ln(\pi/\beta\alpha\lambda)/2$ and $c_2=1/2$. 

\begin{figure}
\centerline{
\psfig{figure=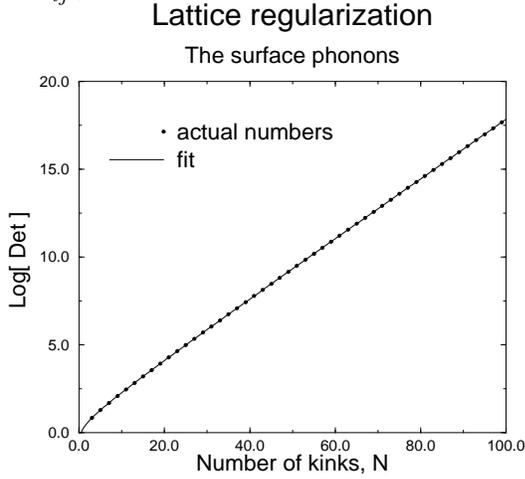,width=3truein}}
{\caption{We calculate numerically the logarithm of the 
surface phonon determinant, $\det M^p_{ij}$,
for $N=2\ ...\ 100$ kinks and fit the result with $f(N)=p_0+p_1 N+ p_2\ln N$
}
\label{f6}}
\end{figure}

We were unable to obtain an analytical expression for  the surface phonon 
determinant $\det M^p_{ij}$. For $N=2...100$ kinks we calculate 
the determinant numerically and fit its logarithm
with $f(N)= p_0+p_1 N+ p_2 \ln N$ (Figure \ref{f6}),
\begin{equation}
\det M^p_{ij}= N^{p_2} \exp\lbrace p_0+p_1 N\rbrace .
\label{169}
\end{equation}
We find $p_0=0.09\pm 0.02$, $p_1=0.166\pm 0.002$ and $p_2=0.24\pm 0.05$. 
(We expect that the surface phonon fluctuations contribute to $Z_f$ 
similar to the curvy cut fluctuations (\ref{168a}) --- hence the 
form of the fitting curve for $\det M^p_{ij}$.)

From (\ref{168}-\ref{169})
\begin{eqnarray}
Z_f&=&\exp\lbrace p_1-p_0\rbrace{\biggl({{32\beta^2\mu\alpha
\lambda^3}\over{\pi^3(1+\chi)}}\biggr)}^{1/2}{\biggl({{\ell_c}\over
\lambda}\biggr)}^{-p_2+1/2}\nonumber\\
&&{\biggl({{\pi^3(1+\chi)\exp\lbrace -2 p_1\rbrace}
\over{32\beta^2\mu\alpha\lambda^3}}\biggr)}^{\ell_c/2\lambda}
\  Z_{\rm continuous}\label{170}
\end{eqnarray}
which following (\ref{120}) gives the imaginary part of the free energy 
in the lattice regularization
\begin{eqnarray}
{\rm Im}F^{\rm lattice}&=&\exp\lbrace p_1-p_0\rbrace
{\biggl({{32\beta^2\mu\alpha
\lambda^3}\over{\pi^3(1+\chi)}}\biggr)}^{1/2}{\biggl({{\ell_c}\over
\lambda}\biggr)}^{-p_2+1/2}\nonumber\\
&&{\biggl({{\pi^3(1+\chi)\exp\lbrace -2 p_1\rbrace}
\over{32\beta^2\mu\alpha\lambda^3}}\biggr)}^{\ell_c/2\lambda}
{\rm Im}F^{\rm simple}\label{171}
\end{eqnarray}
with ${\rm Im}F^{\rm simple}$ from (\ref{i4}).

Throughout the calculation we ignored the kinetic term in the energy of the
elastic material: their behavior is pretty trivial, as momenta and positions
decouple. Because we introduce new degrees of freedom with our ``splitting
atoms'' model for the crack, we discuss the effects of the corresponding new
momenta.  Before ``splitting'', $(\ell_c/\lambda-1)$ atoms along the cut
contribute
\begin{eqnarray}
&&Z_{\rm kinetic}^u=\nonumber\\
&&\prod_{i=1}^{\ell_c/\lambda-1}\int\int dp_{i_x}dp_{i_y}
{\biggl({{\lambda}\over{2\pi\hbar}}\biggr)}^2
\exp\biggl\lbrace-\beta\sum_{j=1}^{\ell_c/\lambda-1}
{{p_{j_x}^2+p_{j_y}^2}\over{2 m}}
\biggr\rbrace\nonumber\\
\label{k1}
\end{eqnarray} 
to the partition function of the uncracked material $Z_0$. 
(We do not consider the contribution
to the partition function from the bulk atoms --- they contribute in 
a same way to $Z_0$ as they do to $Z_1$, and thus drop out from the 
calculation of the imaginary part (\ref{120}).)
The configuration space integration measure for a classical statistical 
system is $dx dp/2\pi\hbar$; because we integrated out the displacements
with the weight $1/\lambda$ to make them dimensionless, 
the momentum integrals have  measure $dp \lambda/2\pi\hbar$, (\ref{k1}).
The formation of the cut increases the number of the kinetic degrees 
of freedom by $(\ell_c/\lambda-1)$ (the number of split atoms).
The split atoms  contribute 
\begin{eqnarray}
Z_{\rm kinetic}^s&=&\prod_{i=1}^{2(\ell_c/\lambda-1)}\int\int dp_{i_x}dp_{i_y}
{\biggl({{\lambda}\over{2\pi\hbar}}\biggr)}^2\nonumber\\
&&\exp\biggl\lbrace
-\beta\sum_{j=1}^{2(\ell_c/\lambda-1)}{{p_{j_x}^2+p_{j_y}^2}\over{ m}}
\biggr\rbrace\label{k2}
\end{eqnarray}
to the partition function of the material with the cut.
From (\ref{120}), (\ref{k1}) and (\ref{k2}) the kinetic energy of 
the elastic material modify the imaginary part of the free energy 
by a factor $Z_k$:
\begin{eqnarray}
{\rm Im}F^{\rm lattice}&\to& Z_k{\rm Im}F^{\rm lattice}  \label{k3}
\end{eqnarray}
with
\begin{equation}
Z_k={{Z_{\rm kinetic}^s}\over{Z_{\rm kinetic}^u}}=
{\biggl({{m\lambda^2}\over{8\pi\beta\hbar^2}}\biggr)}^{\ell_c/\lambda-1}.
\label{k4}
\end{equation}

One might notice that for both ($\zeta$-function and lattice) regularizations 
the effect of the quadratic fluctuations  can be absorbed into the 
renormalization of the prefactor of the imaginary part of the free energy
calculated in a simplified model (without the quadratic fluctuations) 
(\ref{i4}),
and the material surface tension $\alpha$: the multiplicative 
factor to the imaginary  part of the free energy has a generic form
\begin{equation}
{\rm Im} F^{\rm simple}\to n_0{\biggl({{\ell_c}\over\lambda}\biggr)}^{n_1}
\exp\biggl\lbrace n_2 {{\ell_c}\over\lambda}\biggr\rbrace {\rm Im} 
F^{\rm simple}\label{172}
\end{equation}
where the first two terms renormalize 
the prefactor of ${\rm Im} F^{\rm simple}$ and the other one 
can be absorbed into ${\rm Im} F^{\rm simple}$ through the effective 
renormalization of the surface tension
\begin{equation}
\alpha\to\alpha_r=\alpha+{1\over{2\beta\lambda}}n_2
\label{173}
\end{equation}
From (\ref{k3}), (\ref{k4}) it follows that in case of the lattice 
regularization, the inclusion of the kinetic energy 
of the elastic material shifts the constants $n_0$ and $n_2$, thus 
preserving (\ref{172}):
\begin{eqnarray}
n_0&\to& n_0 \biggl({{8\pi\beta\hbar^2}\over{m\lambda^2}}\biggr)\label{k5}\\
n_2&\to& n_2 +\ln{{m\lambda^2}\over{8\pi\beta\hbar^2}}\nonumber
\end{eqnarray} 
The calculation  of the  kinetic terms in the $\zeta$-function regularization
is more complicated. We however have no reason to believe that 
it will change  the form (\ref{172}).

\section{The asymptotic behavior of the inverse bulk modulus}
In our earlier work\cite{we}, we discussed how the thermal instability
of elastic materials with respect to fracture under  infinitesimal
stretching load determines the asymptotic behavior of the high order elastic 
coefficients. Specifically, for the inverse bulk modulus K(P) 
in two dimensions (material under compression) 
\begin{eqnarray}
{1 \over K(P)} &=& -{1 \over A} \left({\partial A \over \partial P}\right)_{
\beta}= c_0 + c_1 P \cdots + c_n P^n + \cdots
\label{179}
\end{eqnarray} 
we found within linear elasticity and ignoring the quadratic 
fluctuations,
\begin{equation}
{c_{n+1} \over c_n}\rightarrow - n^{1/2} 
{\biggl({{ \pi (\chi+1)}\over  {64 \beta\mu \alpha^2 }  } \biggr)}^{1/2}
\mbox{\hspace{0.1in}  as {\it n}} \rightarrow\infty,
\label{180}
\end{equation}
which indicates that the high--order terms $c_n$  roughly grow as $(n/2)!$
and so the perturbative expansion for the inverse bulk modulus 
is an asymptotic one.

In this section we show that, except for the temperature dependent 
renormalization of the surface tension $\alpha\to\alpha_r=\alpha+
O(1/\beta)$, (\ref{180}) remains true even if we include the quadratic 
fluctuations around the saddle point (the critical crack); moreover, 
we argue that (\ref{180}) is also  unchanged by the nonlinear corrections 
to the linear elastic theory near the crack tips.  

\begin{figure}
\centerline{
\psfig{figure=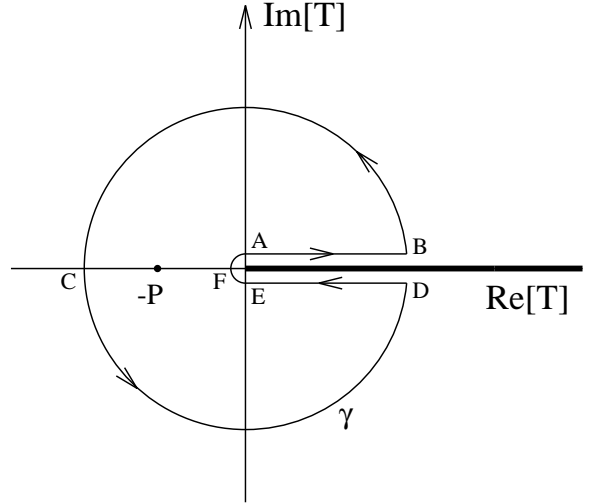,width=3truein}}
{\caption{The free energy of the elastic material $F(T)$ is analytical 
in the complex $T$ plane  except for a branch cut 
$T\in[0,+\infty)$. This allows a Cauchy representation for 
the free energy $F(-P)$ of the material under compression.}
\label{f7}}
\end{figure}

We review how one can calculate the high order coefficients of the 
inverse bulk modulus\cite{we}. The free energy $F(T)$ of the elastic material  
is presumably analytical in the complex $T$ plane function for small
$T$ except for a branch 
cut $T\in[0,+\infty)$ --- the axis of stretching. (We show this explicitly 
in the calculation within  linear elastic theory without the 
quadratic fluctuations\cite{we}.) It is assumed here that neither nonlinear 
effects near the crack tips nor  the quadratic fluctuations 
 change the analyticity domain of the free energy for reasonably 
small $T$ (i.e. $T\leq Y$).
One can then use  Cauchy's theorem to express 
the free energy of the material under compression $F(-P)$, 
(Figure \ref{f7}):   
\begin{equation}
F(-P)={1 \over{2 \pi i}}\oint\limits_{\gamma}{{F(T)\over{T+P}}} dT .
\label{181}
\end{equation} 

The contribution to (\ref{181}) from the arc EFA goes to zero as the 
latter shrinks to a point. In this limit we have
\begin{eqnarray}
F(-P)&=&{1 \over{2 \pi i}}\int\limits_0^B{{F(T+i 0)-F(T-i 0)}\over{T+P}} dT
\nonumber\\
&&+{1 \over{2 \pi i}}\oint\limits_{\rm BCD}{{F(T)}\over{T+P}} dT\nonumber\\
&=&{1 \over{\pi}}\int\limits_0^B{{{\rm Im}F(T)}\over{T+P}} dT
+{1 \over{2 \pi i}}\oint\limits_{\rm BCD}{{F(T)}\over{T+P}} dT .
\label{181a}
\end{eqnarray} 
As it was first established for similar problems in field 
theory\cite{f1} ---\cite{f3}, (\ref{181a}) determines the high--order 
terms in the expansion of the free energy $F(-P)=\sum_n {f_n P^n}$
\begin{equation}
f_n={{(-1)}^n \over \pi}\int\limits_0^B{{{\rm Im} F(T)} \over
{T^{n+1}}} dT+{{(-1)}^n \over {2\pi i}}\oint\limits_{\rm BCD}{{F(T)} \over
{T^{n+1}}} dT .
\label{182}
\end{equation}
The second integral on the RHS of (\ref{182}) produces a
 convergent series; and is hence unimportant to the asymptotics: 
the radius of convergence by the ratio test is of 
the order the radius of the circle BCD (i.e. larger than  $P$ by 
construction). The first integral  generates the asymptotic divergence
of the inverse bulk modulus expansion:
\begin{equation}
f_n\to{{(-1)}^n \over \pi}\int\limits_0^B{{{\rm Im} F(T)} \over
{T^{n+1}}} dT
\mbox{\hspace{0.1in}  as {\it n}} \rightarrow\infty .
\label{182a}
\end{equation}
Once a perturbative expansion for the free energy  is known, one can calculate
the power series expansion for the inverse bulk modulus using  
the thermodynamic relation
\begin{equation}
{1 \over K(P) }={1 \over {P A }}{\biggl ({{\partial F(-P)} \over
{\partial P}}\biggr)_{\beta}}
\label{183}
\end{equation}
so that 
\begin{equation}
{{c_{n+1}}\over{c_n}}={{(n+3)f_{n+3}}\over{(n+2) f_{n+2}}} .
\label{184}
\end{equation}
Note that because the saddle point calculation becomes more and more accurate
as $T \to 0$, and because the integrals in equation (\ref{182a}) are dominated
by small $T$ as $n \to \infty$, using the saddle--point form for the
imaginary part of the free energy yields the correct $n\to\infty$ asymptotic 
behavior of the high-order coefficients $f_n$ in the free energy.
Following (\ref{i4}) and (\ref{172}) the imaginary part of the 
free energy including the quadratic fluctuations is given by
\begin{eqnarray}
{\rm Im} F(T)&=&{{2 n_0} \over {\beta^2  T\lambda^2} }
\biggl(1-{{n_1 n_2}\over{2\beta\lambda\alpha_r}}\biggr)
{\biggl({{2\beta\mu\lambda^2}\over{\chi+1}}\biggr)}^{1/2}
\nonumber\\&&{\biggl(  {{32 \mu
 \alpha_r } \over { \pi T^2 (\chi+1)\lambda}  } \biggr)}^{n_1}
\biggl ({ \pi {{A} \over {\lambda ^2}}} \biggr )\exp{ 
\biggl\lbrace  {{-32 \beta\mu
 \alpha_r^2 } \over { \pi T^2 (\chi+1)}  } \biggr\rbrace}
\nonumber\\ \label{185}
\end{eqnarray}
where $\alpha_r$ is given by (\ref{173}). Note that 
$n_0$, $n_1$, $n_2$ and $\alpha_r$ in (\ref{185}) are  regularization
dependent coefficients, by our calculations in the previous section.
From (\ref{182a}) and (\ref{184}) we find
\begin{equation}
{{c_{n+1}}\over{c_n}}\to -{\biggl({{ \pi (\chi+1)}\over  {32 \beta\mu
 \alpha_r^2 }  } \biggr)}^{1/2}{{(n+3)\Gamma(n/2+n_1+2)}\over
{(n+2)\Gamma(n/2+n_1+3/2)}} .\label{186}
\end{equation}
(In the limit $n\to\infty$ (\ref{186}) is independent of  $B$ 
in (\ref{182a}).) Using
\begin{equation}
{{\Gamma(n/2+n_1+2)}\over{\Gamma(n/2+n_1+3/2)}}\to\sqrt{n\over 2}
\mbox{\hspace{0.1in}  as {\it n}} \rightarrow\infty
\label{187}
\end{equation}
we conclude from (\ref{186}) that
\begin{equation} 
{{c_{n+1}}\over{c_n}}\to -n^{1/2}{\biggl({{ \pi (\chi+1)}\over  {64 \beta\mu
 \alpha_r^2 }  } \biggr)}^{1/2}
\mbox{\hspace{0.1in}  as {\it n}} \rightarrow\infty .
\label{188}
\end{equation}
Equation (\ref{188}) is a very powerful result: it shows that 
apart from the temperature dependent (regularization dependent) 
correction to the surface tension (\ref{173}), the asymptotic ratio of 
the high order  coefficient of the inverse bulk modulus is unchanged
by the inclusion of the quadratic fluctuations (at least for the 
regularizations we have tried).  
One would definitely expect the surface tension to be 
regularization dependent: the energy to break an atomic bond explicitly 
depends on the ultraviolet (short scale) physics, which is excluded in the 
thermodynamic description of the system.   
This has analogies with calculations in field theory, where physical 
quantities calculated in different regularizations give the same 
answer when expressed in terms of the renormalized masses and charges of
the particles\cite{zj}. Here only some physical quantities appear
regularization independent. 

The analysis that leads to (\ref{188}) is based on linear elastic
theory that is known to predict unphysical singularities
near the crack tips. From\cite{m}, the stress tensor component 
$\sigma_{yy}$, for example, has a square root divergence  
\begin{equation}
\sigma_{yy}\sim T\sqrt{{\ell_c}\over{4 r}}, \mbox{\hspace{0.1in}  
as {\it r}} \rightarrow 0
\label{189}
\end{equation}
as one approaches the crack tip. One might expect that the proper 
non-linear description of the crack tips  changes the asymptotic behavior
of the high order elastic coefficients. We argue here that linear 
analysis gives however the correct asymptotic ration (\ref{188}): the 
{\it linear} elastic behavior dominates the {\it nonlinear}
asymptotics within our model.

\begin{figure}
\centerline{
\psfig{figure=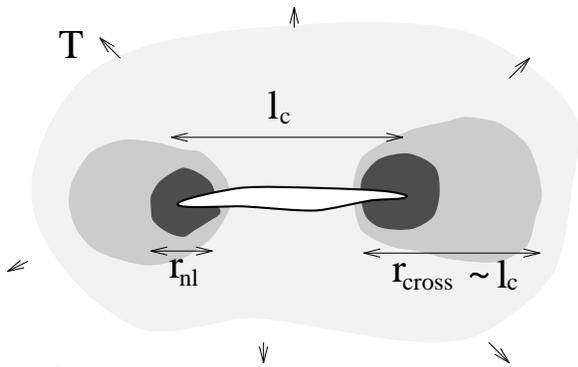,width=3truein}}
{\caption{In the crack system there are two well-defined 
zones: the outer zone, 
consisting of exclusively linear elastic material, and 
the inner, crack tip zone where the nonlinear processes take 
place. Such separation introduces two length scales: the 
first $(r_{\rm nl})$ determines the size of the nonlinear zone, and 
the second $(r_{\rm cross})$ gives  the scale where the elastic 
fields near the crack tip deviate from the ones predicted by SSY to comply 
with the outer zone boundary conditions.}
\label{f8}}
\end{figure}

It is clear that the vital question is how the  energy 
release of the saddle point (critical) crack is changed by  nonlinear 
processes (microcracking, emission of dislocations, etc.) in the vicinity
of the crack tips as $T\to 0$. Following\cite{fbs} we 
distinguish in the crack system two well-defined zones: the outer zone, 
consisting of exclusively linear elastic material, transmits the applied 
traction to the inner, crack tip zone where the nonlinear processes take 
place (Figure \ref{f8}). Such  separation introduces 
two length scales to the problem: $r_{\rm nl}$ and $r_{\rm cross}$. 
The first scale determines the size of the nonlinear process zone near 
the crack tips. It can  be readily 
estimated from (\ref{189}) by requiring the stresses at 
the boundary of the nonlinear zone to be of the order  atomic 
ones $\sigma_{ij}\sim Y$:
\begin{equation}
r_{\rm nl}\sim\ell_c{\biggl({{T}\over {Y}}\biggr)}^2\sim{\alpha\over Y} .
\label{190}
\end{equation}
The second length scale 
is a crossover length $r_{\rm cross}$ where the elastic fields near a 
crack tip deviate from the inner zone $\sqrt{r}$ strain asymptotics 
to depend on the outer-zone boundary conditions (i.e. the length of 
the crack in our case). Normally, $r_{\rm cross}$ is only 
a few times smaller than the crack length\cite{ewalds}, 
\cite{jenif} ---  for  the present calculation 
we assume $r_{\rm cross}\sim \ell_c\sim Y\alpha/T^2$, (\ref{106}).

First, let's consider the energy in the nonlinear zone.
The saddle point energy is $\alpha\ell_c$ and diverges 
as $1/T^2$ as $T\to 0$, while the elastic energy in the  nonlinear zone 
 $E_{\rm nl}$ is bounded by the linear value
\begin{equation}
E_{\rm nl}\sim \int_{0}^{r_{\rm nl}}dr\ r \sigma_{ij}^2(r)/Y
\sim \alpha^2/Y.
\label{191}
\end{equation} 
Since $E_{\rm nl}$ is fixed as $T\to 0$, it renormalizes $n_0$ in 
(\ref{172}) and hence does not affect the asymptotics (\ref{188}).

Second, we consider how the existence of the inner (nonlinear) zone 
changes the energy in the outer (linear) zone. 
The  elastic equations around the crack tip 
allow many solutions\cite{jenif}; in each, the stresses 
$\sigma_{ij}$ have the form  $C_b r^b f_b(\theta)$, $r_{\rm nl}\ll 
r\ll r_{\rm cross}$, in polar coordinates $(r,\theta)$ centered 
at the crack tip, where $b$ is an  half-integer, the $C_b$ are constants, 
and the $f_b$ are known trigonometric functions. 
Linear fracture mechanics predicts  $b=-1/2$ to be the most singular 
solution (compare with (\ref{189})) only because  modes with $b<-1/2$
would give rise to  singular displacements at the crack tip. 
Incorporation of the nonlinear zone $r<r_{\rm nl}$ however,
removes this constraint. In other words,
the nonlinear zone introduces new boundary conditions for  linear 
elasticity solutions, allowing them to be more singular. 
The dominance of $b=-1/2$ solution is known as small scale yielding
 (SSY) approximation. Analyzing the mode III anti-plane shear fracture, 
Hui and Ruina argued\cite{hui} that SSY approximation becomes more and 
more accurate as $\epsilon=r_{\rm nl}/r_{\rm cross}\to 0$. 
(They expect that the same result can be extended for mode I fracture.) 
Clearly, in our case $\epsilon\to 0$  as $T\to 0$; thus 
the dominant contribution still comes from $b=-1/2$ solution.
In fact, following \cite{jenif} we expect 
\begin{eqnarray}
\sigma_{ij}\lbrace C_n\rbrace&=&T\biggl(C_{-1/2}\sqrt{{\ell_c}\over{r}}
\nonumber\\ &&+\sum_{n=\lbrace
\cdots,\ -7/2,\  -5/2,\ -3/2\rbrace}C_n{\biggl({r\over {r_{\rm nl}}}\biggr)}^n
\nonumber\\ &&+\sum_{n=\lbrace \ 1/2,\ 3/2,\ 5/2,\cdots\rbrace} C_n 
{\biggl({r\over {l_c}}\biggr)}^n\biggr).
\label{191a}
\end{eqnarray}
The inelastic stresses at the outer boundary of the nonlinear zone
$r\sim r_{\rm nl}$ are of order $Y$, thus from (\ref{191a}),  for $n<-1/2$, 
$C_n=O(\epsilon^{-1/2})$ 
(recall that $\epsilon=r_{\rm nl}/r_{\rm cross}\sim (T/Y)^2$).
These more singular terms in turn generate corrections to $C_n$
with $n\ge 1/2$ of order $O(\epsilon)$. (One can see this from the fact 
that the dominant contribution from the more singular terms at
$r\sim \ell_{c}$ is $C_{-3/2} ({\ell_c}/r_{\rm nl})^{-3/2}\sim \epsilon$.)  
The dependence of $C_n$ in (\ref{191a}) on the polar angle $\theta$ is implied.

There is a formal analogy between the arguments presented here for 
the stress fields in the crossover zone with the quantum mechanical 
problem of the bound states of the hydrogen atom. 
When we treat hydrogen nucleus as a point charge, for each orbital 
quantum number, the electron wave function has two solutions near 
the origin (the position of the nucleus): one is finite as $r\to 0$ 
and the other one is divergent\cite{ll,ps}.
In a point charge problem one immediately discards the divergent 
solution because it can not be normalized and thus can not 
represent a bound state. However,
in a finite-size nucleus model one notices that the electron wave
function outside the nucleus is a mixture of the finite and the 
divergent solutions of the point charge problem. The normalization 
is resolved because inside the nucleus the electron wave function 
satisfies a different equation and becomes finite. 
The radius of the nucleus serves as a short-distance cutoff similar 
to $r_{\rm nl}$ in the crack problem.      

The  change in the contribution to the saddle point energy 
from the outer zone as a result of the introduction 
of the nonlinear zone, $\delta E_{\rm outer}$, is 
given by
\begin{equation}
\delta E_{\rm outer}\sim \int_{r_{\rm nl}}^{\ell_c} dr\ r{{
\sigma_{ij}^2\lbrace C_n\rbrace-\sigma_{ij}^2\lbrace C_n^{\rm linear}
\rbrace}\over Y}.
\label{192a}
\end{equation}
The dominant contribution to (\ref{192a}) comes from the cross term between 
$n=-1/2$ and $n=-3/2$ corrections in (\ref{191a}):
\begin{equation}
\delta E_{\rm outer}\sim {{\alpha^2}\over Y} \ln{{T\over Y}}:
\label{192}
\end{equation} 
the correction  renormalizes the  $n_1$ 
coefficient in the imaginary part of the  free energy (\ref{185}) 
(regularization dependent in a first place), leaving the asymptotic 
ratio (\ref{188}) intact.

It is no surprise that the nonlinear effects do not change 
the generic form of the imaginary part (\ref{185}). The
detailed nonlinear description of the crack tips is a specification
of the ultraviolet (short scale) physics and thus is nothing but 
another choice of the regularization. From our experience with
$\zeta$-function and the lattice regularizations, we  
naturally expect that this {\it nonlinear} regularization 
preserves the form of the imaginary part (\ref{185}). 

Finally, let's consider the enhanced nucleation of secondary cracks 
in the high-strain outer-zone region --- a possible cause for breakdown
of the ``dilute gas'' approximation. Inside the nonlinear zone of 
the saddle point crack, the critical crack length for a second crack
is of the order $\alpha/Y$ ( from (\ref{106}) with $T\sim Y$ ) and thus such 
microcracks can be easily created. In fact, the nucleation of 
these microcracks may well be the dominant mechanism of the main crack 
propagation. Micro-crack nucleation in the nonlinear zone will
change the stress fields near the crack tips, but as we discuss above, 
has little impact on the saddle point energy (as the total energy in the 
nonlinear zone is finite). We show now that such secondary crack nucleation
is exponentially confined to the nonlinear zone of the main crack.
The probability $W(r_0)$ of the second crack nucleated somewhere at 
$r>r_0\sim r_{\rm nl}$ ($r=0$ corresponds to a crack tip) is given by
\begin{equation}
W(r_0)\sim \int\limits_{r_0}^{+\infty}{{r dr}\over{\lambda^2}}
\exp\lbrace -\beta\alpha\ell(r)\rbrace
\label{195}
\end{equation}  
where $\ell(r)$ is a critical crack length at distance $r$ from the 
tip of the critical crack. From (\ref{106}) with $T$ replaced with 
the stress field near the crack tip given by (\ref{189}), we find
\begin{equation}
\ell(r)\sim {{\alpha Y}\over{{\sigma_{ij}(r)}^2}}
\sim r 
\label{196}
\end{equation}
(\ref{195}) with (\ref{196}) gives
\begin{equation}
W(r_0)\sim \int\limits_{r_0}^{+\infty}{{r dr}\over{\lambda^2}}
\exp\lbrace -\beta\alpha r\rbrace = {{1+\beta\alpha r_0}\over{
{(\beta\alpha\lambda)}^2}}\exp\lbrace -\beta\alpha r_0\rbrace
\label{198}
\end{equation}
The exponential dependence of $W(r_0)$ on the boundary 
of the nonlinear zone $r_0\sim r_{nl}$  in equation (\ref{198})
means that the nucleation of another crack (in addition to the 
saddle point one) is exponentially confined to the nonlinear zone, 
justifying the dilute gas approximation.

\section{Other geometries, stresses, and fracture mechanisms}
In this section we discuss generalizations of our model,
more exactly its simplified version without the quadratic fluctuations. 
We will do five things.
In {\bf (A)} we calculate the imaginary part of the free energy for 
arbitrary uniform loading and find the high-order nonlinear corrections
to  Young's modulus. We discuss the effects of dislocations and
vacancy clusters (voids) in {\bf (B)} and {\bf (C)}. Part {\bf (D)} deals with 
three dimensional fracture through the nucleation of penny-shaped
cracks: we calculate the imaginary part of the free energy 
and the asymptotic ratio of the successive coefficients of the 
inverse bulk modulus. Finally, in {\bf (E)}  we consider a non-perturbative
effect: the vapor pressure of a solid gas of bits fractured from 
the crack surfaces, and show how it affects the saddle point calculation.  

\subsection{Anisotropic uniform stress and the high order corrections 
to Young's modulus.}
We calculated the essential singularity of the free energy 
at zero tension only for  uniform {\it isotropic} loads at infinity. 
Within the approximation of ignoring the quadratic fluctuations,
we can easily generalize to any uniform loading. In general, consider
an infinite elastic material subject to  a uniform asymptotic tension
with $\sigma_{\rm yy}=T$, $\sigma_{\rm xx}=\epsilon T$ ($0\le\epsilon < 1$)
and $\sigma_{\rm xy}=0$.
Using the strain-stress analysis of\cite{m} and following 
(\ref{36})-(\ref{60}), we find the energy $E_{\rm release}$, 
released from the 
creation of the straight cut of length $\ell$ tilted by angle $\theta$ from 
the $x$ axis
\begin{equation}
E_{\rm release}={{\pi T^2\ell^2(1+\chi)}\over{64\mu}}\biggl[
(1+\epsilon)+(1-\epsilon)\cos 2\theta\biggr].
\label{d2}
\end{equation}
(The isotropic result (\ref{st}) is restored for $\epsilon=1$.)
The new important feature that comes into play is that the crack rotation
ceased to be a zero-restoring-force mode. 
Treating the crack rotation to quadratic order in $\theta$ from the 
saddle point value $\theta=0$, we obtain the total energy 
of the crack $E(\Delta\ell,\theta)$  similar to  (\ref{105}-\ref{107}),
(\ref{107b})
\begin{equation}
E(\Delta\ell,\theta)=\alpha\ell_c-{{\alpha\Delta\ell^2}\over{\ell_c}}
+\alpha\ell_c (1-\epsilon)\theta^2.
\label{d4}
\end{equation}
As before, $\Delta\ell$ is the deviation of the crack length 
from the saddle point value $\ell_c$, still given by (\ref{106}).
Following (\ref{i1})-(\ref{120}), the imaginary part of the free energy for a 
dilute gas of straight cuts, excluding all quadratic fluctuations
except for the uniform contraction-expansion  (mode $\Delta\ell$) 
and  the rotation (mode $\theta$) of the critical droplet,  is given by
\begin{eqnarray}
{\rm Im} F^{\rm simple}(T,\epsilon)&=&\pm{\pi \over {2\beta^2\alpha\lambda} }
\biggl ({ {{A} \over {\lambda ^2}}} \biggr ){\biggl({1\over{1-\epsilon}}
\biggr)}^{1/2}\nonumber\\
&&\exp{ \biggl\lbrace  {{-32 \beta\mu
 \alpha^2 } \over { \pi T^2 (\chi+1)}  } \biggr\rbrace}.
\label{d5}
\end{eqnarray} 
One immediately notices an intriguing fact: the $\epsilon$-dependence of 
the imaginary part is only in the prefactor, which, as  we already know 
is  regularization dependent anyway.
In particular, the latter means that the inverse Young's modulus
--- the elastic coefficient corresponding to the transition 
with path $\epsilon=0$ --- will have the same asymptotic behavior 
as that of the inverse bulk modulus (\ref{188}): the asymptotic ratio of the 
high-order elastic coefficients of the inverse Young's modulus $Y(P)$
\begin{eqnarray}
{1 \over Y(P)} &=& -{1 \over A} \left({\partial A \over \partial P}\right)_{
\beta}= Y_0 + Y_1 P \cdots + Y_n P^n + \cdots
\label{d6a}
\end{eqnarray} 
($P$ in (\ref{d6a}) is a  uniaxial compression) are given by 
\begin{equation} 
{{Y_{n+1}}\over{Y_n}}\to -n^{1/2}{\biggl({{ \pi (\chi+1)}\over  {64 \beta\mu
 \alpha^2 }  } \biggr)}^{1/2}
\mbox{\hspace{0.1in}  as {\it n}} \rightarrow\infty.
\label{d6}
\end{equation}

\subsection{Dislocations}
We have forbidden dislocation nucleation and plastic flow in our
model.  Dislocation emission is crucial for ductile fracture, but 
by restricting ourselves to a brittle fracture of defect--free
materials we have escaped many  complications.  
Dislocations are in principle important: the nucleation\cite{nelson}
barrier $E_{\rm dis}$ for two edge dislocations in an isotropic 
linear--elastic material  under a uniform tension $T$ with equal and 
opposite Burger's vectors $\vec b$ is
\begin{equation}
E_{\rm dis}={{Y b^2}\over {4 \pi (1-\sigma^2)} }\ln {Y\over T} + E_0 
\label{d1}
\end{equation}
where $E_0$ is a $T$ independent part that includes the dislocation 
core energy.
The fact that $E_{\rm dis}$ grows like $1/\ln T$ as $T\to 0$ (much more
slowly than the corresponding barrier for cracks) tells that 
in more realistic models dislocations and the resulting plastic 
flow\cite{ambegaokar} cannot be  ignored.
While dislocations may not themselves lead to a catastrophic instability
in the theory (and thus to an imaginary part in the free energy?),
they will strongly affect the dynamics of crack nucleation (e.g., crack 
nucleation on grain boundaries and dislocation tangles)\cite{ewalds,fbs}.

\subsection{Vacancy clusters}
We ignore void formation. It would seem natural to associate the 
negative pressure (tension) $(-T)$ times the unit cell size with the 
chemical potential $\mu$ of a vacancy.  At negative chemical potentials, 
the dominant fracture mechanism becomes the nucleation of vacancy 
clusters or voids (rather than Griffith-type  microcracks), 
as noted by Golubovi\'c and collaborators\cite{golubovic}. 
If we identify the chemical potential of a vacancy with $-T$, 
we find the total energy of creation a circular vacancy of radius $R$,
 $E_{\rm vac}(R)$, to be
\begin{equation}
E_{\rm vac}(R)=2\pi R\alpha- T\pi R^2. 
\label{d7}
\end{equation}
From (\ref{d7}) the radius of the critical vacancy is  $R_c=\alpha/T$ and
its energy is given by $E_{\rm vac}(R_c)=\pi\alpha^2/T$.
A saddle point is a circular void because a circular void gains the most 
energy ($\sim$ area of the void) for a given perimeter length. 
In principle, the exact shape of the critical cluster is also affected 
by the elastic energy release. The latter, however,  
\begin{eqnarray}
E_{\rm release}(R_c)={{\pi T^2 R_c^2 (3\chi+1)}\over{8\mu}}
={{\pi \alpha^2 (3\chi+1)}\over{8\mu}}
\label{d8}
\end{eqnarray}
is fixed as $T\to 0$, and thus the energy of the vacancy is 
dominated by $E_{\rm vac}(R_c)$ for small $T$. 
(To obtain (\ref{d8}) we used the strain-stress analysis of\cite{m}
and expression (\ref{60}) for the energy release.)
Using the framework developed for the  crack nucleation, we find that in 
case of voids (again, ignoring the positive frequency quadratic fluctuations)
the imaginary part of the free energy is given by
\begin{equation}
{\rm Im} F^{\rm simple}_{\rm vacancy}
(T)=\pm{1 \over {2\beta} }
\biggl ({ {{A} \over {\lambda ^2}}} \biggr ){\biggl({1\over{\beta T\lambda^2}}
\biggr)}^{1/2}\exp{ \biggl\lbrace  {{-\pi \beta
 \alpha^2 } \over  T  } \biggr\rbrace}.
\label{d9}
\end{equation}
(The special feature of the calculation (\ref{d9}) is that translations
are the only zero modes: the rotation of a circular vacancy cluster 
does not represent a new state of the system.)
From (\ref{d9}) we obtain following (\ref{182}) and (\ref{184})
the asymptotic ratio of the high-order coefficients of the inverse
bulk modulus 
\begin{equation}
{{c_{n+1}}\over{c_n}}\to -{n\over {\pi \beta \alpha^2}}.
\label{d10}
\end{equation}
The divergence of the inverse bulk modulus is much stronger in 
this case: the high-order coefficients grow as $c_n\sim n!$, rather
than as $(n/2)!$  (for the fracture through the crack nucleation).

Whether (\ref{d10}) is a realistic result is  an open 
question. Fracture through  vacancy cluster nucleation is 
an unlikely mechanism for highly brittle materials: 
the identification of $\mu$ with $(-T)$ demands a mechanism for 
relieving elastic tension by the creation of vacancies. 
The only bulk mechanism for vacancy formation is
dislocation climb, which must be excluded  from consideration ---
the dislocations in highly brittle materials are immobile\cite{fbs}.
Vacancy clusters might be important for the fracture of ductile (non-brittle) 
materials. However, the nucleation of  vacancies must be considered 
in parallel with the nucleation of dislocations. Because at small $T$
dislocations are nucleated much more easily (\ref{d1}) than vacancy clusters 
at low stresses, the dominant bulk mode of failure is much more likely to 
be crack nucleation at a dislocation tangle or grain boundary --- as 
indeed is observed in practice.

\subsection{Three dimensional fracture}
Our theory can be extended to describe a three dimensional fracture
transition as well. Studying elliptical cuts, Sih and Liebowitz\cite{sl} 
found that a penny-shaped 
cut in a three-dimensional elastic media subject to a uniform isotropic 
tension $T$ relieves the most elastic energy for a given area of the cut.
The energy to create a penny-shaped cut of radius $R$, $E_{\rm penny}(R)$,
is given by\cite{sl}  
\begin{equation}
E_{\rm penny}(R)=2\pi\alpha R^2 - {{4(1-\sigma)R^3 T^2}\over{3\mu}}.
\label{d11}
\end{equation} 
The zero modes contribute in this case a factor $2\pi V/\lambda^3$ ---
$2\pi$ coming from the distinct rotations of the cut, 
and $V/\lambda^3$ coming 
from the translations of the cut.
Here we find the imaginary part of the free energy to be
\begin{eqnarray}
{\rm Im} F^{\rm simple}_{\rm penny}
(T)&=&\pm{1 \over {2\beta} }
\biggl ({2\pi {V \over {\lambda ^3}}} \biggr )
{\biggl({1\over{2\beta \alpha\lambda^2}}
\biggr)}^{1/2}\nonumber\\
&&\exp{ \biggl\lbrace  {{- 2\beta\mu^2\pi^3\alpha^2
  } \over  {{3 (1-\sigma)}^2 T^4 } } \biggr\rbrace}
\label{d12}
\end{eqnarray}
and the asymptotic ratio of the high-order elastic coefficients of the 
inverse bulk modulus
\begin{equation}
{{c_{n+1}}\over{c_n}}\to -{\biggl({{{3 (1-\sigma)}^2
}\over{2\beta\mu^2\pi^3\alpha^2}}\biggr)}^{1/4}{\biggl({n\over 4}
\biggr)}^{1/4}.\label{d13}
\end{equation}

\subsection{Vapor pressure}
The approach we used to calculate the imaginary part of 
the free energy is a perturbative one. In a sense, nothing prohibits 
us from considering cubic, quartic, etc. deviations from the saddle
point energy. In fact, it is possible to develop an analog of 
Feynman diagram technique (as in quantum electrodynamics\cite{ll} or 
quantum field theory\cite{zj}) and calculate 
the contribution to the imaginary part to  any finite order.
It is  important to realize that even if we did this, the result would
still be incomplete: we would miss  interesting and important 
physics coming from  nonperturbative effects.
Here we discuss one such nonperturbative effect, namely the ``vapor'' 
pressure of a solid gas of bits fractured from the crack surface. 
We find that including the vapor pressure, the essential singularity 
shifts from $T=0$ to $T=-P_{\rm vapor}$.
Consider a dilute gas of straight cuts of arbitrary length  with an 
elliptical opening (mode $v_1$ in (\ref{z6})) and a solid gas of 
fractured bits from the crack surface. Following\cite{we}, the
partition function of the material with one cut $Z_1$ under 
a uniform isotropic tension $T$ is 
\begin{eqnarray}
Z_1 &=& Z_0\biggl(\pi{A\over{\lambda^2}}\biggr)\int_{0}^{
\infty}{{d \ell}\over{\lambda}}\int_{0}^{\infty}
{{dv_1}\over{\lambda}} \exp\biggl\lbrace -\beta \biggl(
2\alpha \ell\label{vp1}\\
&&+{{2\pi \mu}\over {\chi+1}}v_1^2 -{{\pi T\ell}\over{2}} v_1\biggr)
 \biggr\rbrace Z_{\rm gas}(\ell,v_1)\nonumber,
\end{eqnarray}  
where $Z_{\rm gas}(\ell,v_1)$ is the partition function 
of the gas of fractured 
bits inside the crack of area $\pi\ell v_1/2$. It costs $2\alpha\lambda$ 
to fracture one bit of size $\lambda\times  \lambda$ from a  crack 
step, so in an ideal gas approximation, the partition
function of the  gas is determined by
\begin{eqnarray}
Z_{\rm gas}(\ell,v_1)&=&\sum_{n=0}^{\infty}\exp(-2\beta\alpha\lambda N)
{\biggl({{\pi\ell v_1}\over{2\lambda^2}}\biggr)}^N {1\over {N!}}
\label{vp2}\\
&=&\exp\biggl\lbrace {{\pi\ell v_1}\over{2\lambda^2}}\exp
(-2\beta\alpha\lambda)\biggr\rbrace.\nonumber
\end{eqnarray} 
From (\ref{vp1}) and (\ref{vp2}) it follows that the partition 
function of the  gas effectively increases the tension $T$ by 
the vapor pressure $P_{\rm vapor}$, $T\to T+P_{\rm vapor}$, where
\begin{equation}
P_{\rm vapor}={1\over{\beta\lambda^2}}\exp(-2\beta\alpha\lambda).
\label{vp3}
\end{equation}
In particular, the essential singularity 
of the free energy shifts from zero tension to minus the ``vapor''
pressure. This shift is clearly a nonperturbative 
effect. We were able to describe it only by allowing topologically 
different excitations in the system: a state of the elastic 
material  with  a bit completely detached from the crack surface 
{\it may not} be obtained by the continuous 
deformation of the crack surface (surface phonons) or the cut shape
(curvy cuts). 
At zero  external pressure, our material is in the gas (fractured) phase
--- not until $P_{\rm vapor}$ is the solid stable.

\section{Summary}
In this paper we studied the stress-induced phase transition of 
elastic materials under external stress: an elastic compression 
``phase'' under positive pressure goes to a fractured ``phase''
under tension. We showed that in a properly formulated
thermodynamic limit, the free energy of an infinite elastic  material 
with holes of predetermined shapes is independent of the shape of the 
outer boundary as the latter goes to infinity. Under a
stretching load the free energy develops an imaginary part with an essential
singularity at  vanishing tension. To calculate the essential singularity
of the free energy including  quadratic fluctuations we 
determined the spectrum and  normal modes of 
surface fluctuations of a straight cut, and  proved that 
under the uniform isotropic tension a 
curvy cut releases the same elastic energy (to cubic order) as the 
straight one with the same end-points. 
The imaginary part of the free energy determines the asymptotic 
behavior of the high-order nonlinear correction to the 
inverse bulk modulus\cite{we}. We find that although the prefactor 
and the renormalization of the surface tension are both regularization 
dependent (once we include the quadratic fluctuations), the 
asymptotic ratio of the high-order successive coefficients of the inverse 
bulk modulus apparently is a regularization-independent result.

Within our model, the asymptotic ratio is unchanged by the inclusion 
of the nonlinear effects near the crack tips.
We generalize the simplified model (without the quadratic fluctuations)
to anisotropic uniform strain and calculated the asymptotic behavior 
of the high order nonlinear coefficients of the inverse Young's 
modulus. We computed the imaginary part of the free energy
(and the corresponding divergence of the high-order coefficients of 
the inverse bulk modulus)  for the fracture mechanism through void 
nucleation which dominates at small external pressures: we argue 
that it may not occur in brittle fracture and should be preempted by
dislocation motion in ductile fracture.
We find that the simplified model applied to  three-dimensional 
fracture predicts a $(n/4)!$ divergence of the nonlinear coefficients 
of the inverse bulk modulus. 

Our results can be viewed as a straightforward extension to the
solid-gas sublimation point of Langer \cite{{langer},{langer2}} and
Fisher's \cite{fisher} theory of the essential singularities at the
liquid-gas transition.  Indeed, if we allow for vapor pressure in our
model, then our system will be in the gas phase at $P=0$, 
as noted in section VII(E).  The essential
singularity we calculate shifts from $P=0$ to the vapor pressure.
If we measure the nonlinear bulk modulus as an expansion about (say)
atmospheric pressure, it should converge --- but the radius of convergence
would be bounded by the difference between the point of expansion and the
vapor pressure.

\section*{Acknowledgment}
We acknowledge the support of DOE Grant DE-FG02-88-ER45364. 
We would like to thank Yakov Kanter, Eugene Kolomeisky, Paul Houle, Tony 
Ingraffea, Paul Wawrzynek, Lisa Wickham, Herbert Hui, Ken Burton and 
Robb Thompson for useful conversations.


\begin{references}

\bibitem{g} A. A. Griffith, {\it Philos. Trans. Roy. Soc. London
Ser. A} {\bf 221} 163 (1920).

\bibitem{inglis} C. E. Inglis, {\it Trans. Inst. Naval Archit.}
{\bf 55} 219 (1913).

\bibitem{fisher} M. E. Fisher, {\it Physics} {\bf 3} 255 (1967).
Recent work has shown that some first-order phase transitions 
can have power-law singularities:  D.~J. Bukman and J.~J.~D. Shore,
{\it J.~Stat. Phys.} {\bf 78} 1277 (1995).

\bibitem{selinger} R. I. B Selinger, Z. Wang and W. M. Gelbart, 
{\it Phys. Rev. A} {\bf 43} 4396 (1990) and references therein.

\bibitem{k1} R. L. Smith, S. L. Phoenix, M. R. Greenfield,
R. B. Henstenburg and R. E. Pitt {\it Proc. R. Soc. Lond. A}
{\bf 388} 353 (1983) and references therein; also Ken Burton,
unpublished.

\bibitem{dyson} F. J. Dyson {\it Phys. Rev.} {\bf 85} 631 (1952). 

\bibitem{zj} J. Zinn-Justin, ``Quantum field theory and critical phenomena'',
 Oxford University Press, New York ( 1989).

\bibitem{b} H. F. Bueckner, {\it J. App. Mech., Trans. ASME} {\bf 80}
1225 (1958).

\bibitem{rice2} J. R. Rice, ``Mathematical analysis in the mechanics 
of fracture'', in {\it Fracture} (ed. H. Liebowitz) vol. {\bf 2}, 
Academic Press, New York (1968). 

\bibitem{good} J. N. Goodier, ``Mathematical theory of equilibrium
cracks'', in {\it Fracture} (ed. H. Liebowitz) vol. {\bf 2}, 
Academic Press, New York (1968). 

\bibitem{sl} G. C. Sih and H. Liebowitz, ``Mathematical theories 
of brittle fracture'', in {\it Fracture} (ed. H. Liebowitz) vol. {\bf 2}, 
Academic Press, New York (1968). 


\bibitem{g1} N. Goldenfeld, ``Lectures on phase transitions and the 
renormalization group'', Addison-Wesley Publishing Company, 
Advanced Book Program (1993).

\bibitem{we} A. Buchel and J. P. Sethna, {\it Phys. Rev. Lett.}
{\bf 77} 1520 (1996). Here, calculating the imaginary part of the free
 energy we incorrectly found the contribution from the zero-restoring-force
modes. Because straight cuts tilted by $\theta$ and $\pi+\theta$ with respect
to, say, the $X$ axis are identical, the correct contribution from rotations
should be $\pi$, rather than $2\pi$. This constant does not change the other 
results of the paper.

\bibitem{ewalds} H. L. Ewalds and R. J. H. Wanhill, ``Fracture 
mechanics'', Edward Arnold and Delftse Uitgevers Maatschappij,
Delft (1984). 

\bibitem{rice} J. R. Rice, {\it J. App. Mech., Trans. ASME} {\bf 35}
379 (1968).

\bibitem{betti} I. S. Sokolnikoff, ``Mathematical theory of
elasticity'', McGraw-Hill Book Company, New York (1956).

\bibitem{m} N. I. Muskhelishvili, ``Some basic problems of the
mathematical theory of elasticity; fundamental equations, plane theory
of elasticity, torsion and bending'',
Groningen, P. Noordhoff (1963).

\bibitem{langer} J. S. Langer, {\it Ann. Phys.} {\bf 41} 108 (1967);
see also N. J. G\"unther, D. A. Nicole and D. J. Wallace, 
{\it J. Phys. A} {\bf 13} 1755 (1980).


\bibitem{langer2} J. S. Langer, {\it Ann. Phys.} {\bf 54} 258 (1969).

\bibitem{affleck} I. Affleck, {\it Phys. Rev. Lett.} {\bf 46} 388 (1981).

\bibitem{ramond}  P. Ramond, ``Field theory : a modern primer'',
Reading, Mass. : Benjamin/Cummings Pub. Co., Advanced Book
Program (1981).

\bibitem{rushic} I.S. Gradshteyn and I.M. Ryzhik, ``Table of integrals, 
series, and products'', Academic Press, Boston (1994).

\bibitem{f1} C. M. Bender and T. T. Wu, Phys. Rev.
 {\bf 184}, 1231 (1969).

\bibitem{f2} E. Bre\'zin, J. C. Le Guillou and J. Zinn-Justin, 
Phys. Rev. D {\bf 15} 1558 (1977).

\bibitem{f3} G. Parisi, {\it Phys. Lett.} B {\bf 66} 167 (1977).

\bibitem{fbs} B. R. Lawn and T. R. Wilshaw, ``Fracture of brittle solids'',
Cambridge University Press, Cambridge (1975).

\bibitem{jenif} J. A. Hodgdon and J. P. Sethna, {\it Phys. Rev.} 
B {\bf 47} 4831 (1993); J. A. Hodgdon, ``Three dimensional fracture : 
symmetry and stability'', PhD thesis, (1993). 

\bibitem{hui} C. Y. Hui and Andy Ruina, {\it Int. J. Frac.} {\bf 72}
97 (1995). They consider  both  integer and half-integer $r^n$ corrections
to the stress fields near the crack tips.
Hodgdon and Sethna\cite{jenif} found that only half-integer corrections 
are needed to describe the influence of the nonlinear zone.
 Hui and Ruina used an over-complete set of 
functions to describe the angular dependence of the stress fields. 

\bibitem{ll} A. I. Akhiezer and V. B.
Berestetski, ``Quantum electrodynamics'',  Interscience Publishers,
New York (1965).

\bibitem{ps} I. Pomeranchuk and S. Smorodinsky, {\it J. Phys.} {\bf 9}
97 (1945).

\bibitem{nelson} D. R. Nelson, {\it Phys. Rev.} B {\bf 18} 2318 (1978).

\bibitem{ambegaokar} Two-dimensional dislocation--mediated plastic flow is 
closely related to the problem of vortex nucleation and superflow decay in 
superfluid films: V.~Ambegaokar, B.~I. Halperin, D.~R. Nelson, and 
E.~D. Siggia, {\it Phys. Rev. Lett.} {\bf 40}, 783 (1978);
V.~Ambegaokar, B.~I. Halperin, D.~R. Nelson, and E.~D. Siggia,
{\it Phys. Rev.} B {\bf 21}, 1806 (1980); P.~Minnhagen, O.~Wetman,
A.~Jonsson, and P.~Olsson, {\it Phys. Rev. Lett.} {\bf 74}, 3672 (1995).

\bibitem{golubovic} L. Golubovi\'c  and A. Peredera, {\it Phys. Rev.} E
{\bf 51} 2799 (1995).  They conclude with the interesting and apparently
correct statement that voids smaller than the Griffith length can
nonetheless grow and fracture the material.


\end{references}
\end{document}